\documentclass[aps,pra,showpacs,amsmath,amssymb,preprintnumbers,superscriptaddress,10pt,twocolumn]{revtex4-1}

\usepackage{amsmath,amssymb}
\usepackage{bm}% bold math
\usepackage[latin1]{inputenc}
\usepackage{graphicx}% include figure files
\usepackage[mediumqspace]{SIunits}
\usepackage{hyphenat}
\usepackage[bookmarks=false]{hyperref}
\usepackage{comment}
\usepackage{color}
\usepackage{multirow}
\usepackage{subcaption}

\definecolor{teal}{RGB}{0,127,127}

\begin{document}

\title{Attacks exploiting deviation of mean photon number in quantum key distribution\\ and coin tossing}

\author{Shihan~Sajeed}
\email{ssajeed@uwaterloo.ca}
\affiliation{Institute for Quantum Computing, University of Waterloo, Waterloo, ON, N2L~3G1 Canada}
\affiliation{Department of Electrical and Computer Engineering, University of Waterloo, Waterloo, ON, N2L~3G1 Canada}

\author{Igor~Radchenko}
\affiliation{General Physics Institute, Russian Academy of Sciences, Moscow, 119991 Russia}

\author{Sarah~Kaiser}
\affiliation{Institute for Quantum Computing, University of Waterloo, Waterloo, ON, N2L~3G1 Canada}
\affiliation{Department of Physics and Astronomy, University of Waterloo, Waterloo, ON, N2L~3G1 Canada}

\author{Jean-Philippe~Bourgoin}
\affiliation{Institute for Quantum Computing, University of Waterloo, Waterloo, ON, N2L~3G1 Canada}
\affiliation{Department of Physics and Astronomy, University of Waterloo, Waterloo, ON, N2L~3G1 Canada}

\author{Anna Pappa}
\affiliation{Department of Physics and Astronomy, University College London, London WC1E 6BT, United Kingdom}

\author{Laurent~Monat}
\affiliation{ID~Quantique SA, Chemin de la Marbrerie~3, 1227 Carouge, Geneva, Switzerland}

\author{Matthieu~Legr{\' e}}
\affiliation{ID~Quantique SA, Chemin de la Marbrerie~3, 1227 Carouge, Geneva, Switzerland}

\author{Vadim~Makarov}
\affiliation{Institute for Quantum Computing, University of Waterloo, Waterloo, ON, N2L~3G1 Canada}
\affiliation{Department of Physics and Astronomy, University of Waterloo, Waterloo, ON, N2L~3G1 Canada}
\affiliation{Department of Electrical and Computer Engineering, University of Waterloo, Waterloo, ON, N2L~3G1 Canada}

\date{\today}

\begin{abstract}
The security of quantum communication using a weak coherent source requires an accurate knowledge of the source's mean photon number. Finite calibration precision or an active manipulation by an attacker may cause the actual emitted photon number to deviate from the known value. We model effects of this deviation on the security of three quantum communication protocols: the Bennett-Brassard 1984 (BB84) quantum key distribution (QKD) protocol without decoy states, Scarani-Ac{\' i}n-Ribordy-Gisin 2004 (SARG04) QKD protocol, and a coin-tossing protocol. For QKD, we model both a strong attack using technology possible in principle, and a realistic attack bounded by today's technology. To maintain the mean photon number in two-way systems, such as plug-and-play and relativistic quantum cryptography schemes, bright pulse energy incoming from the communication channel must be monitored. Implementation of a monitoring detector has largely been ignored so far, except for ID~Quantique's commercial QKD system Clavis2. We scrutinize this implementation for security problems, and show that designing a hack-proof pulse-energy-measuring detector is far from trivial. Indeed the first implementation has three serious flaws confirmed experimentally, each of which may be exploited in a cleverly constructed Trojan-horse attack. We discuss requirements for a loophole-free implementation of the monitoring detector.
\end{abstract}

\maketitle

\section{Introduction}
\label{sec:intro}

Since the proposal of the Bennett-Brassard 1984 (BB84) protocol \cite{bennett1984}, there has been much interest in the feasibility of secure quantum key distribution (QKD). A number of security proofs have been proposed \cite{mayers1996,lo1999, shor2000, lutkenhaus2000, renner2005} and successful implementations were carried out \cite{bennett1992b,schmitt-manderbach2007, stucki2009}. However, device models used in the security proofs have often differed from the properties and behavior of the actual equipment, which opened exploitable security loopholes \cite{vakhitov2001,gisin2006,makarov2006,lo2007,zhao2008,lydersen2010a,sun2011}. Most successful attacks are followed by either a physical countermeasure, a modified QKD protocol, or a modified security proof incorporating the imperfection of the device into the model \cite{gottesman2004,fung2009}. Thus, looking for inconsistencies between the devices and their models in the security proof has a high impact on the security verification of the QKD systems. 
 
In previous studies of QKD employing weak coherent pulses, Alice chose the optimum value of her mean photon number $\mu$ based on the line loss, to maximize the secure key rate \cite{lutkenhaus1999,lutkenhaus2000,hwang2003,ma2005}. However, in this work we consider the case when the actual $\mu$ emitted by Alice is larger than this optimum value without Alice knowing this. This can happen because of an active manipulation by Eve, or because Alice underestimates $\mu$ owing to a finite precision of her calibration. We explore the bound on the information that Eve can gain by exploiting this. We also pinpoint imperfections in an existing commercial QKD system that allows Eve to actively change $\mu$, before we introduce the theory. However readers only interested in the attack theory may now skip to Sec.~\ref{sec:theory}.

\nocite{stucki2002,radchenko2014.LaserPhysLett-11-065203,pappa2014.NatCommun-5-3717,kim2015} %fixing reference order because of early inclusion of figure source code
\begin{figure*}
\includegraphics[width=0.95\textwidth]{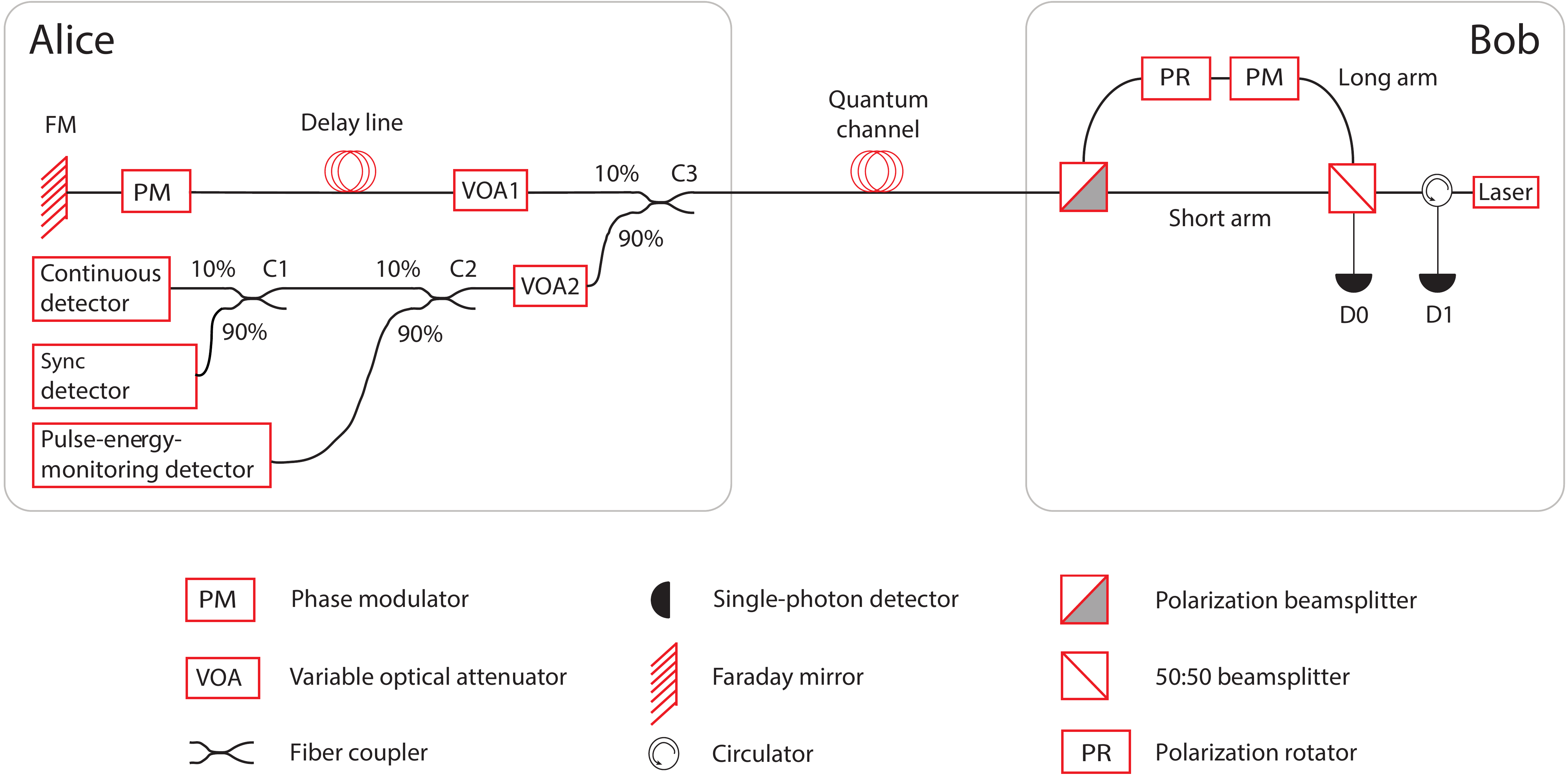}
\caption{(Color online) Plug-and-play system, as implemented in Clavis2 \cite{stucki2002,idqclavis2specs}.} 
\label{fig:plug_and_play}
\end{figure*}

This security issue and our theory is applicable to any QKD scheme that uses weak coherent states. However it is especially important for two-pass schemes. Two-pass optical schemes have significant practical advantages and are widely used today, e.g.,\ in plug-and-play QKD \cite{stucki2002}, relativistic quantum cryptography \cite{radchenko2014.LaserPhysLett-11-065203}, coin-flipping \cite{pappa2014.NatCommun-5-3717}, and most recently to simplify implementation of a measurement-device-independent QKD \cite{kim2015}. In any two-pass scheme, it is necessary for security to monitor the light coming to Alice from Bob (or to Alice and Bob from Charlie, in case of the measurement-device-independent QKD). Otherwise, Eve could substitute a brighter pulse and check the reflected signal to estimate the bit value sent by Alice \cite{vakhitov2001,gisin2006}. Implementation of the monitoring detector has largely been ignored in experimental realizations so far. The first implementation has been done in ID~Quantique's commercial QKD device Clavis2 \cite{idqclavis2specs}, which we describe in Sec.~\ref{sec:description}. We then show in Sec.~\ref{sec:hacking} that the current implementation of the monitoring detector is incapable of being perfectly secure. We demonstrate three flaws in its electronic circuit and show experimentally that each of these flaws can be exploited to compromise the security. Theoretical modeling in Secs.~\ref{sec:theory} and \ref{sec:performance} confirms that even a practical attack implementable today would breach security of this implementation. We develop a general theory of attacks that exploit a changed $\mu$. Section~\ref{sec:theory} proposes both a strong attack that is possible in principle but not currently implementable, and the practical attack that uses off-the-shelf components. Section~\ref{sec:performance} plots performance of the attacks for a range of system parameters. In Sec.~\ref{sec:coin-tossing}, we discuss the applicability of our attacks to the case of practical quantum coin-tossing. We discuss how to redesign the pulse-energy-monitoring detector in a secure way in Sec.~\ref{sec:countermeasures}, and conclude in Sec.~\ref{sec:conclusion}.

\section{QKD system under test}  
\label{sec:description}

\subsection{Plug-and-play scheme}
\label{sec:plug_and_play}

Most fiber-based implementations of QKD systems use either photon polarization or phase encoding of the bit values. However, keeping the polarization stable over long distances in fiber is difficult due to fiber's birefringence that effectively applies a random, time-varying unitary transformation on the polarization state of the photons. To avoid this difficulty, a phase-based plug-and-play QKD system was proposed \cite{muller1997}. As this scheme is implemented in the Clavis2 system (Fig.~\ref{fig:plug_and_play}), we will summarize it here.

The pulses originate in Bob's laser at a rate of $5\,\mega\hertz$ (one pulse every $200\,\nano\second$) and, after passing through an unbalanced Mach-Zehnder interferometer (MZI), they go into the quantum channel. For each pulse generated by the laser there are two orthogonally polarized pulses in the optical link with a delay of $50\,\nano\second$, as the path difference between the two arms of Bob's interferometer is $10\,\meter$. The second pulse has lower energy than the first pulse because it came through the longer arm consisting of the phase modulator (in off state during the first pass), which caused additional loss. At Alice, these bright pulses encounter a 10:90 coupler C3. Only 10\% of the light is used for QKD while the rest is used for synchronization and security purposes. Alice's attenuator VOA1 provides desired attenuation, her phase modulator (PM) applies random phase $\phi_A (0, \frac{\pi}{2}, \pi, \frac{3\pi}{2}$) on the second pulse, and the Faraday mirror (FM) reflects both pulses and rotates their polarization orthogonally. The two pulses, having arrived at Bob, take the opposite arms of the MZI than the ones they took before. The PM in the long arm now applies a random phase $\phi_B$ (either $0~or~\frac{\pi}{2}$). As a result of the combination of FM and unbalanced MZI, the two pulses have the same polarization, path difference and arrive at Bob's 50:50 beamsplitter (BS) at the same time. Hence, the choice of the output BS path depends only on their relative phase difference ($\phi =  \phi_A - \phi_B$). Two detectors $D_0$, $D_1$ and a circulator are used in the configuration shown in the Fig.~\ref{fig:plug_and_play} to collect the light after the BS. If $\phi = 0~(\phi = \pi$), the pulses emerge at the same~(different) path from which they came, and are collected by $D_1$ ($D_0$). However, if Alice and Bob choose different bases (such that $\phi = \frac{\pi}{2}$ or $\frac{3\pi}{2}$), then the photons are split with equal probability between $D_0$ and $D_1$.

\begin{figure*}
\begin{subfigure}[t]{.73\columnwidth}
\includegraphics[width=\columnwidth]{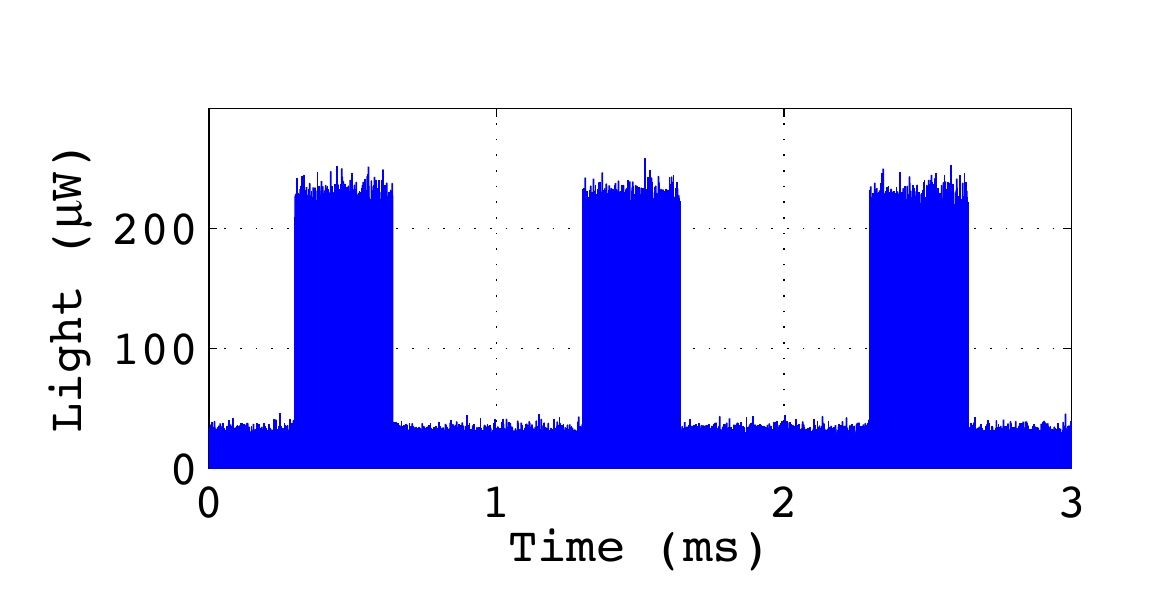}
\caption{}
\label{fig:train}
\end{subfigure}
\begin{subfigure}[t]{.66\columnwidth}
\includegraphics[width=\columnwidth]{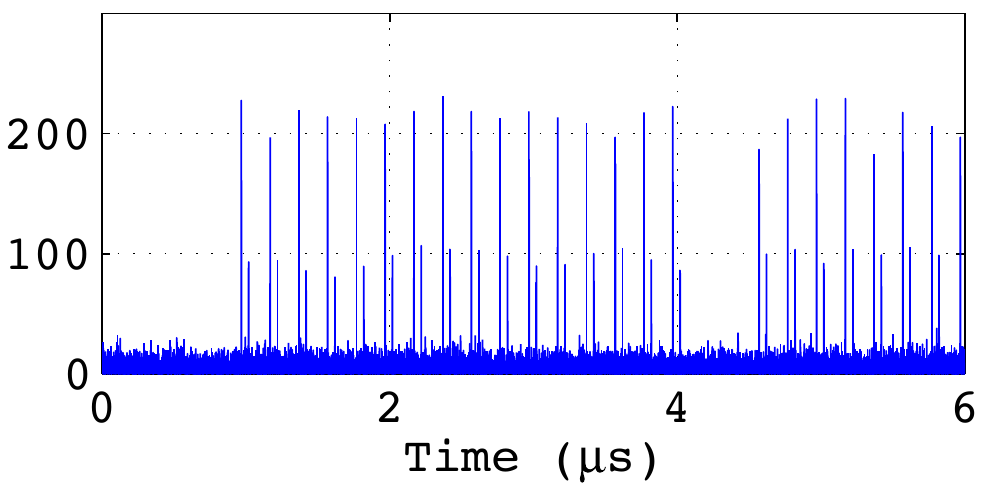}
\caption{}
\label{fig:sync_pulses}
\end{subfigure}
\begin{subfigure}[t]{.66\columnwidth}
\includegraphics[width=\columnwidth]{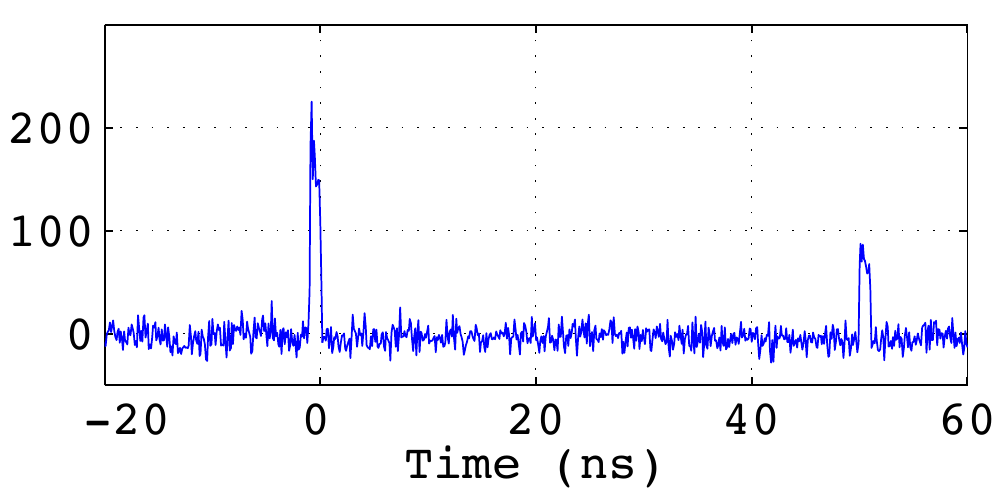}
\caption{}
\label{fig:pulse_energy}
\end{subfigure}
\caption{(Color online) Optical pulses coming into Alice. (a) Trains of pulses (frames) generated by Bob. The frames are generated every $1\,\milli\second$, are $340\,\micro\second$ long, and contain 1700 pulse pairs. (b) Beginning of the frame showing a synchronization pattern. The synchronization circuit checks for this specific pattern in every frame. (c) Two pulses per slot in the optical link. The energy of the first pulse is measured to be $150\,\femto\joule$ and the energy of the second (`calibrated signal pulse') is measured to be $73\,\femto\joule$.}
\label{fig:synchronization}
\end{figure*}

At the input of Alice, $\approx 90\%$ of the incoming light is split at C3 towards the continuous, sync and pulse-energy-monitoring classical detectors (Fig.~\ref{fig:plug_and_play}). A variable attenuator VOA2 is intended to be used for complementing the channel loss to provide a constant amount of power to these detectors. In the rest of this section, we elaborate some technical aspects of system operation that the reader needs to know before we could explain our hacking.

\subsection{Synchronization}
\label{sec:sync}

The synchronization of Alice's clock to Bob's clock is provided by the sync detector (Fig.~\ref{fig:plug_and_play}). The synchronization is required for QKD, and must thus be maintained under any successful attack. Pulses are coming from Bob in packets called \textit{frames} generated every $1\,\milli\second$ as shown in Fig.~\ref{fig:train}. Each frame is $340\,\micro\second$ long and contains 1700 pulse pairs with $200\,\nano\second$ period. Each of the $200\,\nano\second$ intervals containing one of these pulse pairs is called a \textit{slot.} 

In Clavis2, only the first 20 slots of each frame are used for the synchronization of Alice's clock,  i.e.,\ the timing of Alice's modulator to Bob's laser modulator and detector. If they are detected as expected, the particular frame is considered to be synchronized. The synchronization pulses are shown in Fig.~\ref{fig:sync_pulses}. In the beginning of each frame, Bob first sends 16 pulses, then skips two pulses ($17^{th}$ and $18^{th}$) intentionally, and then sends the rest of the pulses of the frame. Alice's synchronization detector checks for this pattern in the first $20$ slots with an avalanche photodiode receiver (Fujitsu FRM5W232BS). Upon detection of the correct pattern, Alice's electronics clock is synchronised to the frame, and further signals from the sync detector are disregarded.  This synchronization is done separately for each frame coming from Bob.

As mentioned in Sec.~\ref{sec:plug_and_play}, for each laser pulse generated per slot there are two signals in the optical link (see Fig.~\ref{fig:pulse_energy}). The energy of the first pulse is measured to be $150\,\femto\joule$ while the energy of the calibrated signal pulse is measured to be $73\,\femto\joule$ at the output of Bob (energy values in the rest of the paper were measured at the same point; Alice--Bob line attenuation was close to $0\,\deci\bel$ in our tests). Note that, in Alice, as the random phase was applied only at the second pulse, only this pulse contains the quantum information and hence we will call it `calibrated signal pulse' for the rest of the paper.

\subsection{Countermeasure against Trojan-horse attack}
\label{sec:countermeasure-design}

In the absence of industry standards for QKD security and prior secure implementations of the plug-and-play scheme, ID~Quantique had to blaze the trail and define an internal standard for implementing and testing the pulse-energy-monitoring detector. First, the implementation had to be compact and inexpensive in order to fit into the rackmounted commercial system. Second, for each system leaving the factory it was to be precisely calibrated and tested against $0.1\,\deci\bel$ ($\approx 2\%$) increase in an energy of a single individual pulse incoming to Alice, the increase being applied {\em on top of} the normal pulse sequence expected at Alice's input. The monitoring detector had to reliably raise alarm in this condition. The testing equipment and software were developed at the factory.

Further testing by an independent hacking team has confirmed that the system passes this factory-defined specification. However it has also revealed that the specification followed by ID~Quantique is too weak and that one can set up other classes of attacks that exploit overlooked security flaws in the present countermeasure implementation. This emphasises the need for open industrial standards and independent certification labs for QKD implementation security.

\subsection{Pulse energy monitoring}
\label{sec:pulse_energy_monitoring}

\begin{figure*}
\begin{subfigure}{.8\textwidth}
\includegraphics[width=\textwidth]{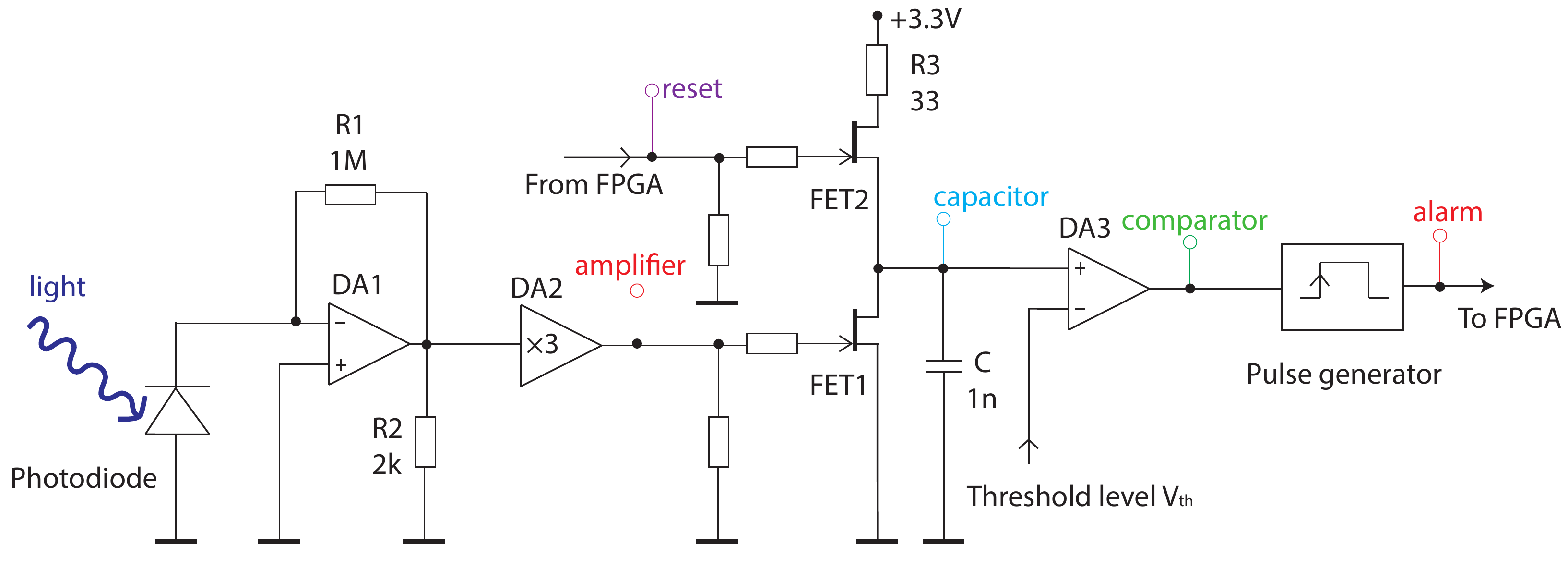}
\caption{Simplified circuit diagram of the pulse-energy-monitoring detector. See text for details.}
\label{fig:schematic}
\end{subfigure}
\begin{subfigure}{\columnwidth}
\includegraphics[width=\columnwidth]{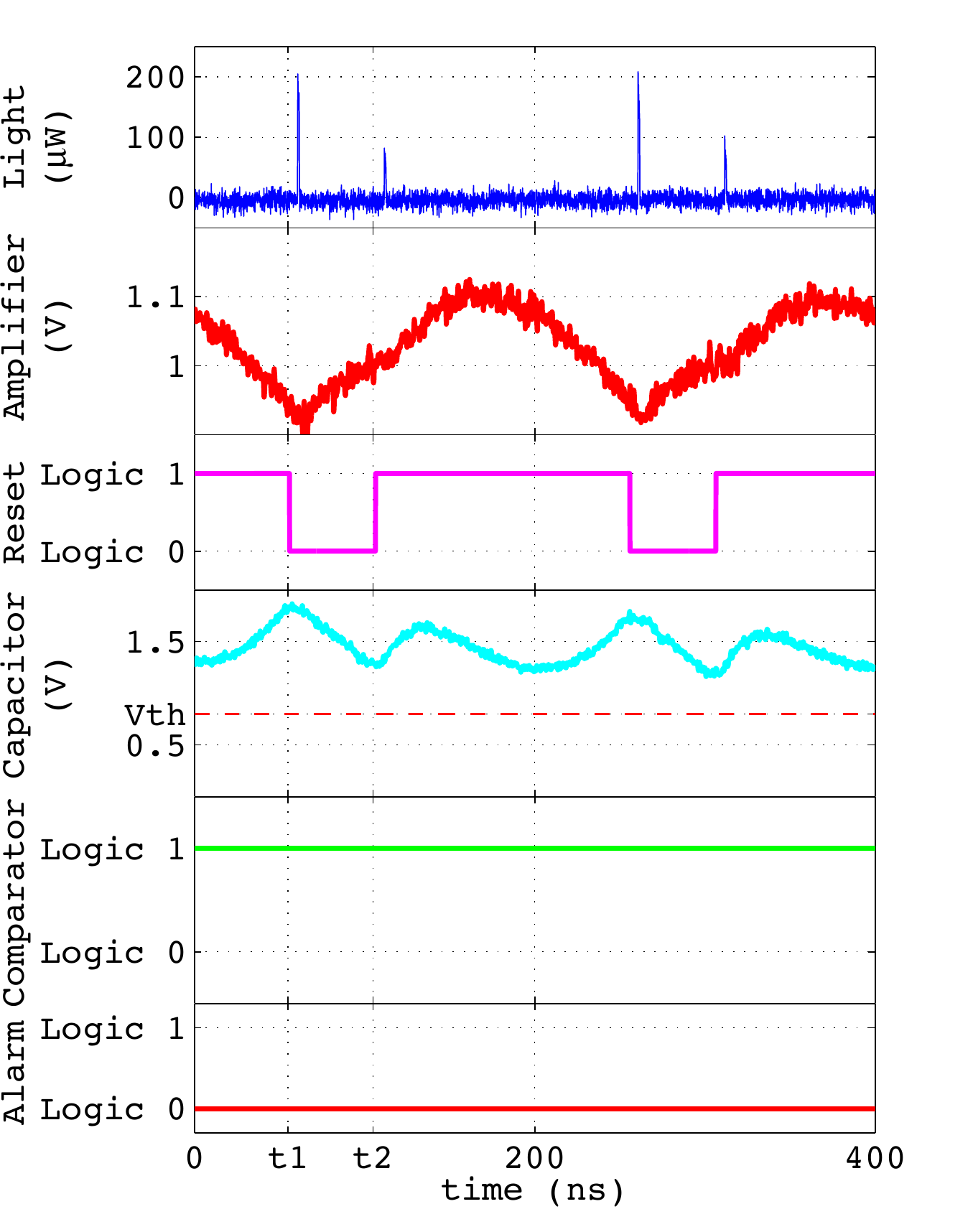}
\caption{Signals during normal operations.}
\label{fig:normal_frame}
\end{subfigure}
\begin{subfigure}{\columnwidth}
\includegraphics[width=\columnwidth]{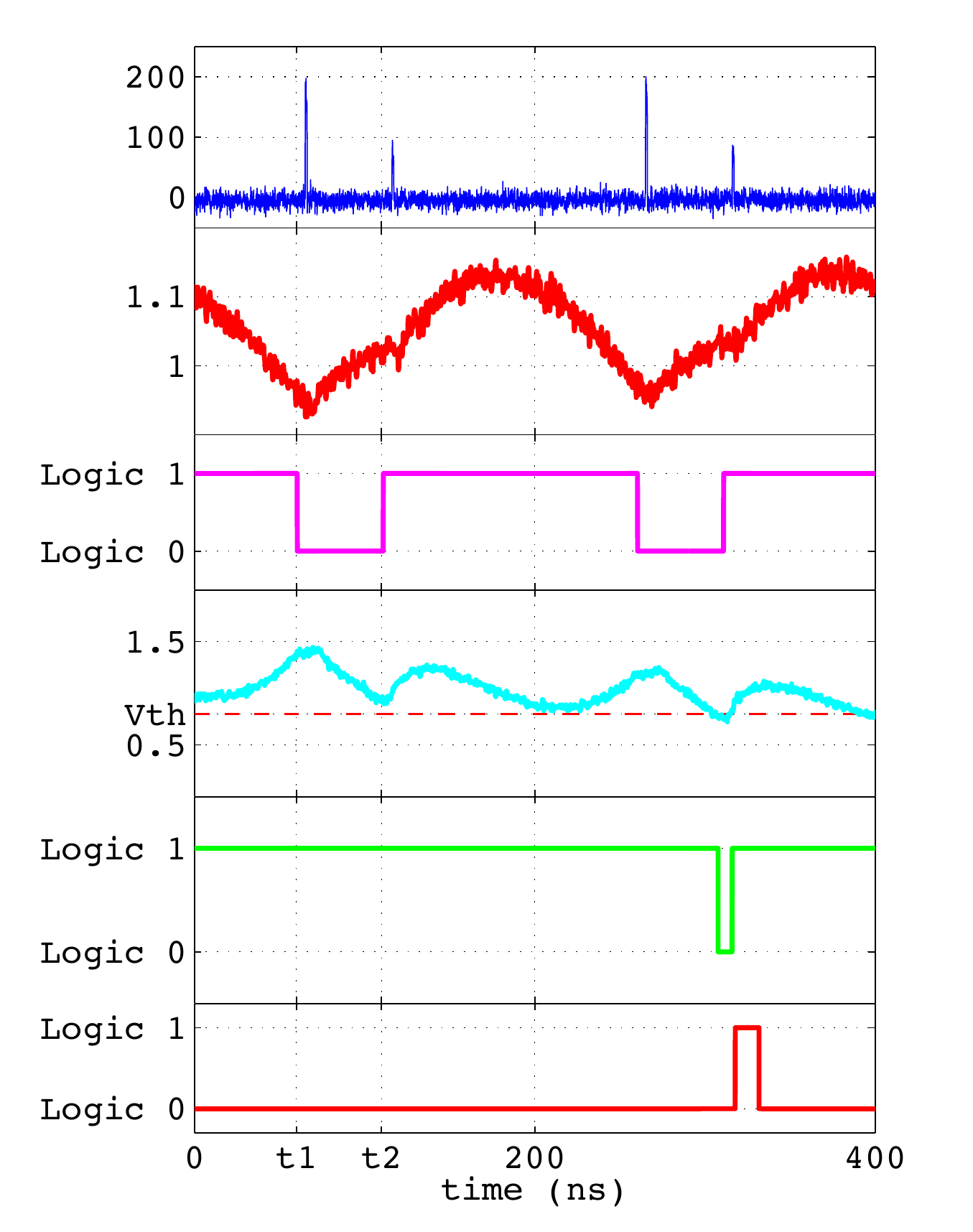}
\caption{Generation of an alert signal on injection of excess light. }
\label{fig:alert_frame}
\end{subfigure}
\caption{(Color online) Pulse-energy-monitoring circuit and oscillograms. The six test points are marked `light', `amplifier', `reset', `capacitor', `comparator' and `alarm' in (a), and the oscillograms at these points are shown (b) for normal operation and (c) for the case when light power is increased by $0.1\,\deci\bel$ (i.e.,\ by $\approx 2\%$) above normal operation. During normal operation, when light pulses arrive with expected energy, the capacitor voltage always stays over the threshold level $V_\text{th}$.  However, when the pulse's energy is higher than expected, due to higher gate pulse to FET1, deeper discharge of the integrating capacitor results. This causes its voltage go below $V_\text{th}$, which in turn creates an alarm.}
\label{fig:pulse-energy-monitoring}
\end{figure*}

\begin{figure*}
\begin{subfigure}[t]{.73\columnwidth}
\includegraphics[width=\columnwidth]{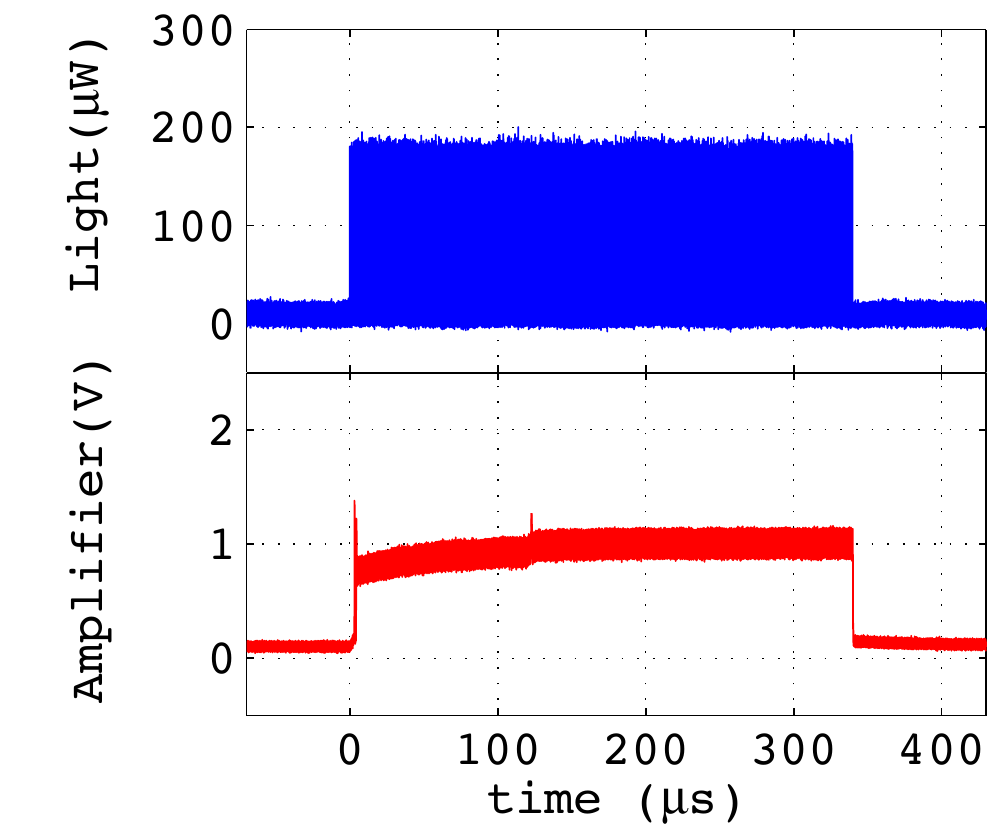}
\caption{}
\label{fig:bump_whole}
\end{subfigure}
\begin{subfigure}[t]{.635\columnwidth}
\includegraphics[width=\columnwidth]{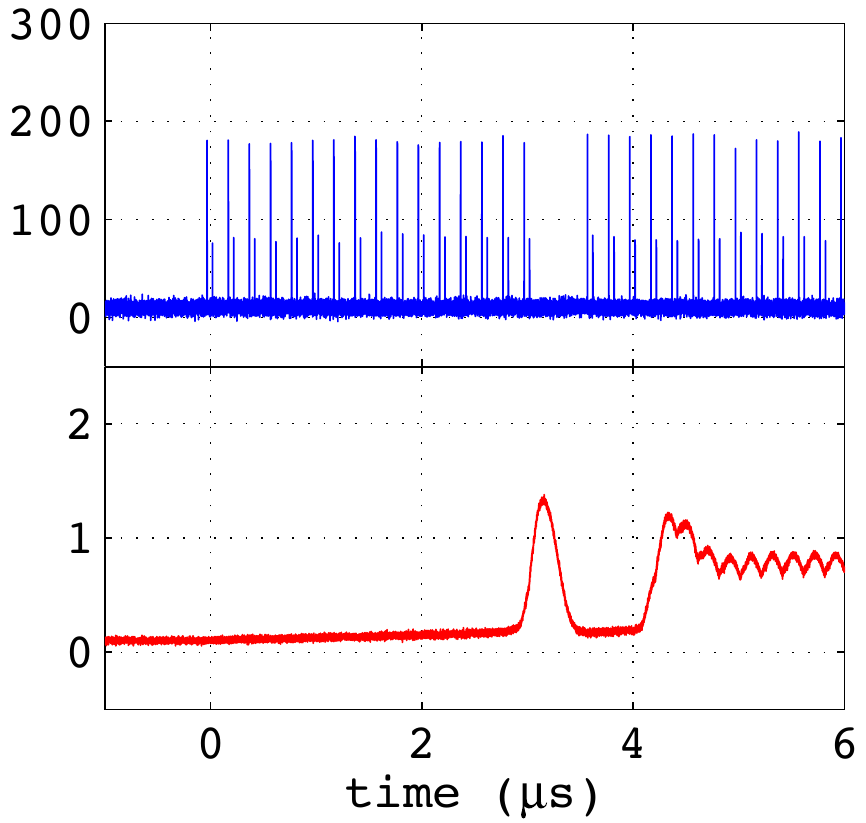}
\caption{}
\label{fig:bump_init}
\end{subfigure}
\begin{subfigure}[t]{.68\columnwidth}
\includegraphics[width=\columnwidth]{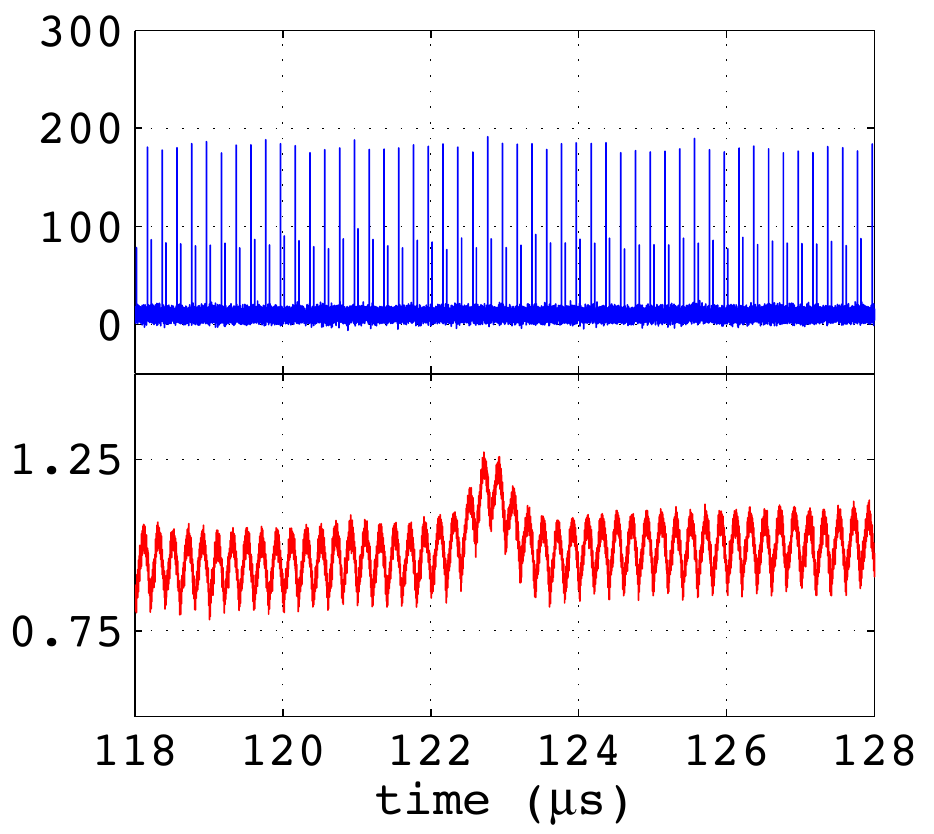}
\caption{}
\label{fig:bump_end}
\end{subfigure}
\caption{(Color online) Recovery of the front-end amplifier from the negative saturation to the normal operation. (a) Entire $340\,\micro\second$ long frame. A minor peak is visible in the amplifier output at $\sim 123\,\micro\second$, marking the recovery of the amplifier from the negative saturation. (b) Initial part of the recovery from the negative saturation. Even though light pulses are arriving at the input of the amplifier, no output is produced for $\sim 3\,\micro\second$. (c) A transient at $\sim 123\,\micro\second$ is the last irregularity, after which the amplifier fully recovers from the saturation.}
\label{fig:bump}
\end{figure*}

The pulse-energy-monitoring circuit is designed to integrate the incoming pulse energy and trigger an alarm when the energy exceeds a predefined threshold value. A fiber-pigtailed p-i-n photodiode~(JDSU EPM~605LL) is used to detect the light. Its photocurrent is processed by an electronic circuit shown in Fig.~\ref{fig:schematic}. Signals at six test points marked in the circuit are shown in Figs.~\ref{fig:normal_frame} and~\ref{fig:alert_frame}. At the front-end of the circuit there is a two-stage transimpedance amplifier, converting photocurrent into voltage signal. Owing to insufficient bandwidth of the amplifier first stage (opamp DA1; Texas Instruments OPA380), it outputs slow-rising electrical pulses that extend to the next few bit slots and interfere with the signals from those slots. The amplifier's second stage is a wideband current-feedback opamp DA2 (Analog Devices AD8009) that does not further distort the signal. Its output acts as a gate pulse for an N-channel field-effect transistor FET1 that is a part of an integrator circuit.

In theory, the operation of the integrator circuit should be the following. The gate pulse for FET2 (reset signal) is applied by the field-programmable-gate-array (FPGA) system controller. This reset signal is normally high, keeping FET2 in a conductive state such that current flows through it to an integrating capacitor C. At time $t_1$, the reset signal switches FET2 into high-impedance state for $50\,\nano\second$, and the capacitor starts to discharge through FET1 (see capacitor signal). The amount of discharge is higher when the power of incoming light is higher. At time $t_2$, reset signal switches FET2 into conductive state again and stops the discharging. This happens in each bit slot, and a negative spike proportional to the incoming light energy is generated at the capacitor. The negative spike is compared to a predefined threshold level $V_\text{th}$, whose value is calibrated at the factory in such a way that during normal operation, the negative spike amplitude is very close but almost never goes below $V_\text{th}$. However, when there is an extra light, this voltage crosses the threshold causing the output of comparator DA3 to go low. 

In actual operation of the practical implementation, when the reset signal from the FPGA goes into the normal high state while the amplifier output is high, both FET1 and FET2 are in the conductive state simultaneously. Instead of charging the capacitor, current from the $+3.3\,\volt$ supply flows through both of them into the ground. As a result, the integrating capacitor cannot be charged instantly by the reset signal. This produces the capacitor signal seen on the oscillogram that does not quite match the expected ideal circuit behavior. Nevertheless, the capacitor signal's lowest level during the cycle strongly depends on the light energy, allowing the circuit to detect a small excessive amount of light in a single pulse when tested to ID~Quantique's specification.

The comparator signal is fed to a pulse generator that produces fixed-width pulse on the low-to-high logic level transition. This is the alarm signal fed to the FPGA that indicates the excess of incoming light. The system software discards all detections in the frame if one or more pulses inside the frame have triggered alarm in Alice. Thus any attempt by Eve to inject brighter pulses in a frame should lead to that frame being dropped from QKD.

\subsection{Frame structure}
\label{sec:data_pulses}

As explained in Sec.~\ref{sec:sync}, the first 20 slots in each frame bear the synchronization pattern. However, data-carrying pulses (which we will henceforth call {\em data pulses}) start from slot 701 ($140\,\micro\second$) and continue to the last slot 1700 ($340\,\micro\second$) of the frame. The slots $21$ to $700$ are idle. The latter is a work-around for an engineering mistake: The output of opamp DA1 enters negative saturation when there is no light coming in \cite{OPA380}. Once pulses appear, recovery from this saturation state takes a relatively long time, approximately $123\,\micro\second$ or 615 slots, with a bump at the end of the recovery (see Fig.~\ref{fig:bump}). Pulse energy alarm signal is only monitored during the data pulses (slots 701--1700) \footnote{We have taken this frame structure from ID~Quantique's factory calibration utility for their commercial encryption products. The current version of QKD software distributed with the research system Clavis2 (as of December 2014) does not perform pulse energy monitoring, and uses a frame structure without the idle pulses.}.

\subsection{Continuous detector}
\label{sec:frame_energy}

The continuous detector in Fig.~\ref{fig:plug_and_play} is low-bandwidth (of the order of $20\,\kilo\hertz$) and is not designed to monitor Bob's individual pulses. The purpose of this detector is to automate the measurement of the line loss at the time of system installation. In addition, this detector may be used for detection of continuous-wave light injected during the Trojan-horse attack, however this functionality has not been implemented and we have not tested it.

\section{Hacking}
\label{sec:hacking}

A general idea of the Trojan-horse attack is that Eve replaces at least some of the data pulses coming from Bob to Alice with brighter ones. These pulses will come out of Alice with proportionally higher mean photon number $\mu$, allowing Eve to exploit their multi-photon statistics to learn more information than expected by Alice and Bob (as will be detailed in Sec.~\ref{sec:theory}). It suffices for Eve to inject only a few bright pulses per frame, because she can exploit these and block all the other Alice's pulses from reaching Bob. 

For a successful attack, Eve must satisfy the following requirements: she must not break the synchronization between Alice and Bob, she must not trigger any alarm, and she must not alter Bob's original detection rate. The total energy of the frame was not monitored in Clavis2 for security (although we also matched it in some of our attacks). To keep the synchronization, we generate pulses in the first 20 slots as expected by the sync detector (see section \ref{sec:sync}). Since in each frame, no monitoring is performed prior to $140\,\micro\second$ (700 pulses; see section \ref{sec:data_pulses}), we are free to generate pulses with any energy till this time to adjust the energy of the whole frame. After pulse energy monitoring begins at $140\,\micro\second$, injecting extra light should trigger an alarm. However, in this section we show three different approaches allowing Eve to inject extra light into at least some calibrated signal pulses without triggering the alarm.

\subsection{Exploiting low bandwidth of front-end amplifier}
\label{sec:BW_hacking}

The $3\,\deci\bel$ bandwidth of the front-end amplifier in the current configuration is about $1\,\mega\hertz$ \cite{OPA380}, which causes it to output slowly rising electrical pulses (as mentioned in Sec.~\ref{sec:pulse_energy_monitoring}). This opens up a loophole which we have experimentally exploited to break the security. We began by redistributing energy between the two pulses of a pair incoming to Alice (shown in Fig.~\ref{fig:pulse_energy}). We suppressed the first pulse and made the second pulse proportionally brighter. Since only the second pulse is modulated in Alice, only its photon number $\mu$ is significant for the security \footnote{This has been shown to be incorrect \cite{ferenczi2012}, however the current Clavis2 software assumes $\mu$ is the mean photon number of the second pulse, disregards the mean photon number of the first pulse, and performs QKD according to these assumptions.}. However because of the slow response of the front-end amplifier, responses to the two pulses largely overlap at the amplifier output. The electronics is thus mainly monitoring the total energy of the pulse pair and not the second pulse. By this method we obtained second pulse energy increase over the calibrated value by a multiplication factor $x = 3.1$, without triggering an alarm. This would break security in theory, but is only sufficient for a partial information leak of 49\% with BB84 protocol (80\% with Scarani-Ac{\' i}n-Ribordy-Gisin 2004 (SARG04) protocol) when using an attack implementable with today's technology (analysed in Secs.~\ref{sec:theory} and \ref{sec:performance}). To increase $x$ further, we then started to suppress additional pulses.

\begin{figure}
\begin{subfigure}[t]{1\columnwidth}
\includegraphics[width=.83\columnwidth]{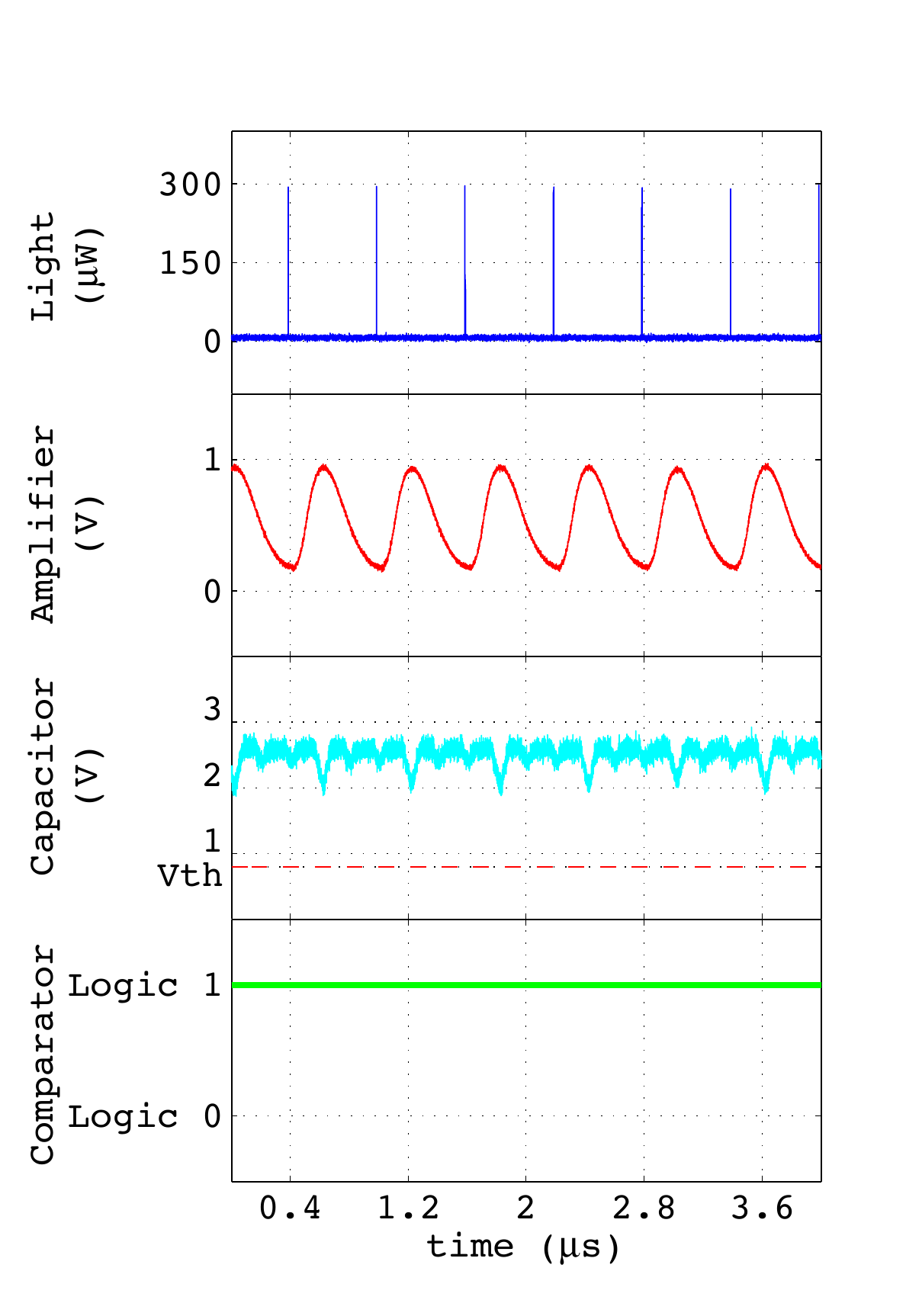}
\caption{Suppression of three pulses out of four and the corresponding effect on the amplifier, capacitor and comparator output. The pulse energy has been increased $8.5$ times from $73\,\femto\joule$ to $623\,\femto\joule$.}
\label{fig:BW_traces}
\end{subfigure}
\begin{subfigure}[t]{.72\columnwidth}
\includegraphics[width=\columnwidth]{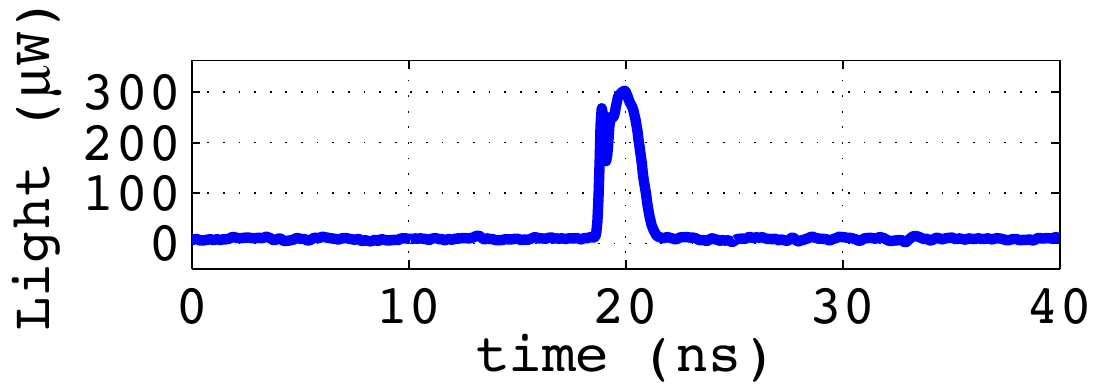}
\caption{Pulse shape carrying the maximum injected energy using this method ($623\,\femto\joule$).}
\label{fig:BW_pulse}
\end{subfigure}
\caption{(Color online) Exploiting the low bandwidth of the front-end amplifier to break the security. }
\label{fig:BW_whole}
\end{figure}

For every four pulses, we suppressed the first three and injected at the fourth slot a bright pulse which we call the `probe pulse' (see Fig.~\ref{fig:BW_whole}). Due to the three blocked pulses, the voltage level at the output of the front-end amplifier is most of the time lower than normal (compare Fig.~\ref{fig:BW_traces} with Fig.~\ref{fig:normal_frame}). When the much brighter probe pulse arrives at the fourth slot, it does not increase the voltage enough to trigger the alarm. In our experiment we were able to inject a probe pulse with a maximum energy of $623\,\femto\joule$  (shown in Fig.~\ref{fig:BW_pulse}), which is approximately $8.5$ times more than the calibrated signal pulse energy ($73\,\femto\joule$).

We also experimentally performed blocking two out of three and one out of two pulses and were able to inject a probe pulse with 7.3 and 5.4 times more energy respectively. We could block more than three pulses but in that case the negative saturation of the amplifier became the dominant factor, as discussed and generalized in the next subsection.

\subsection{Exploiting saturation of front-end amplifier}
\label{sec:saturation_hacking}

\begin{figure}
\includegraphics[width=\columnwidth]{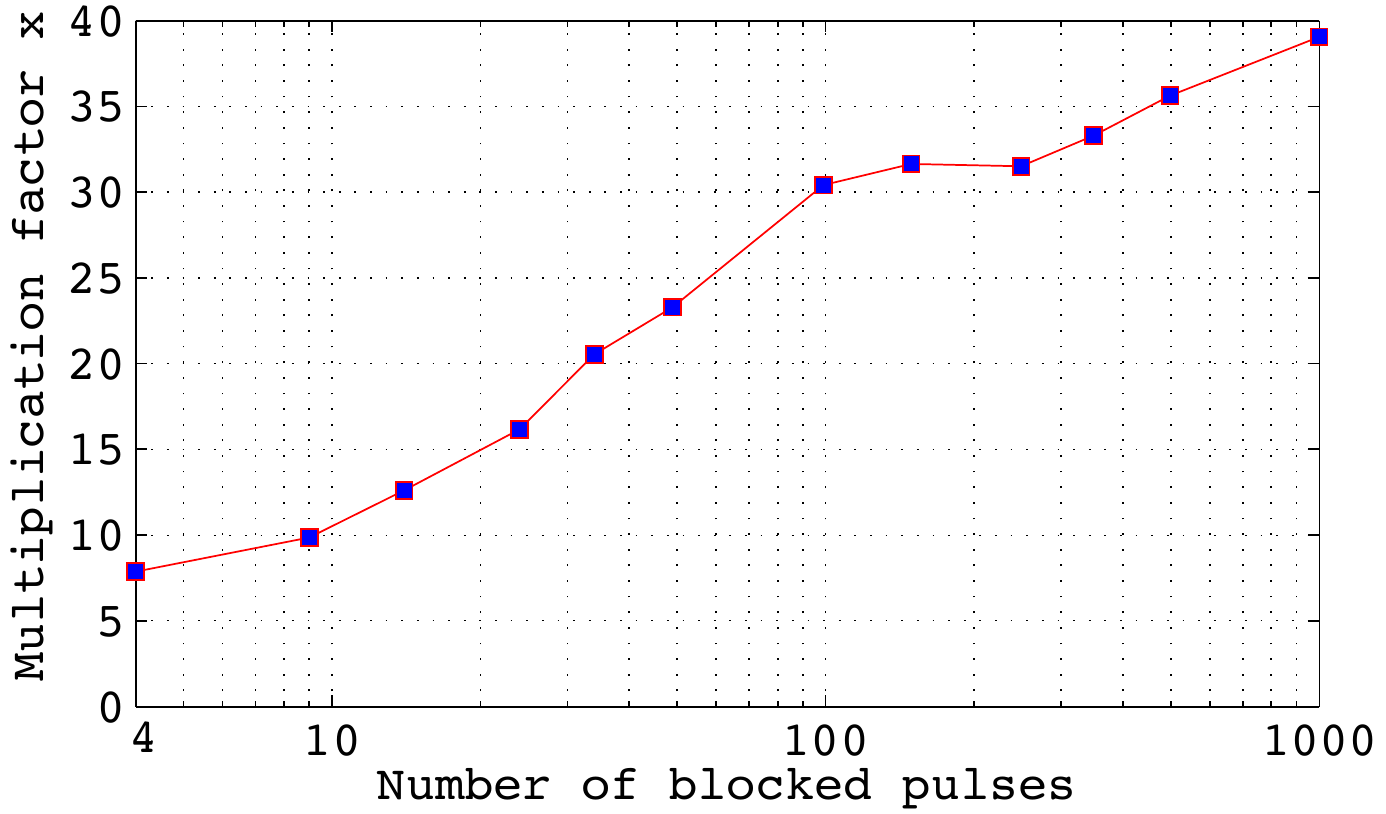}
\caption{(Color online) Energy multiplication factor $x$ for $(n+1)$st pulse vs.\ number of blocked pulses $n$, in the amplifier saturation attack.}
\label{fig:delay}
\end{figure}

As mentioned in section \ref{sec:data_pulses}, data pulses are sent only after $140\,\micro\second$ from the start of the frame because the front-end amplifier takes time to recover from the negative saturation. We removed all the pulses from $100\,\micro\second$ till the start of the monitoring period ($140\,\micro\second$), which forced the amplifier to re-enter the negative saturation. Then, starting at $140\,\micro\second$, for every $n+1$ pulses, we blocked the first $n$ pulses and sent a bright probe pulse at $(n+1)$st slot. We continued to increase the energy of this probe pulse until an alarm was generated. The multiplication factor achieved versus $n$ is plotted in Fig.~\ref{fig:delay}. We see that the curve rises steeply for up to $100$ pulses blocked, then starts to saturate. By blocking $250$ pulses, Eve can achieve multiplication factor $x = 31.5$, while by blocking $100$ pulses she can have $x = 30.4$. Thus, to avoid a reduction of the key rate under attack, it is likely more efficient to block $100$ or fewer pulses. 

\begin{figure}
\includegraphics[width=\columnwidth]{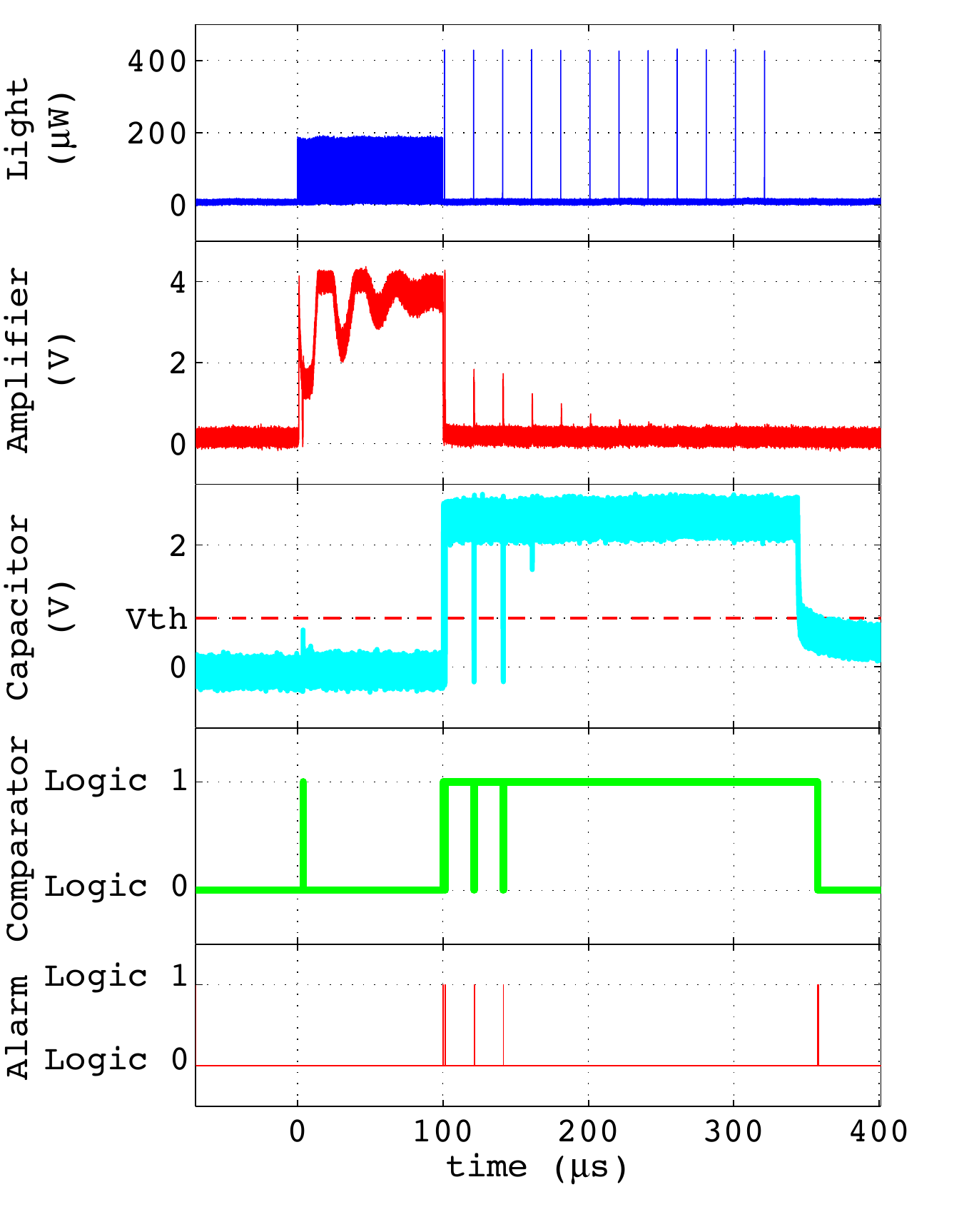}
\caption{(Color online) Attack exploiting the saturation effect of the front-end amplifier, blocking $100$ pulses. The further into the frame the probe pulses are injected, the smaller the amplifier output becomes, because the amplifier stays into saturation for a longer period and more energy is required to bring it out of it. In the alarm plot, the first three pulses occurred because the energy of the probe (light) pulses was enough to produce an amplifier output strong enough to result in an alarm (as it has not yet been into a deep saturation). However, they occurred before the monitoring period and were not counted as an alarm signal by the FPGA. Similarly, the last pulse in the alarm plot occurred when the integrator was reset after the frame (after the end of monitoring period) and was not counted as an alarm.}
\label{fig:saturation_hacking}
\end{figure}

As an example, we show the $100$ pulse blocking case in Fig.~\ref{fig:saturation_hacking}.  Starting from $100\,\micro\second$ into the frame, we began blocking $100$ pulses and sending a bright probe pulse at each $101$st slot. The signal at the amplifier output became smaller as we went further into the frame, vanishing in the last part of it. This is because the longer the amplifier stayed into saturation, the more energy it needed to recover. While we have entered $9$ probe pulses each with $2220\,\femto\joule$ energy ($x = 30.4$), no alarm was generated during the $140$--$340\,\micro\second$ monitoring period.

The reduced pulse rate by itself is not a problem (the maximum number of possible detections in Clavis2 is $19$ per frame owing to the $10\,\micro\second$ deadtime introduced after each detection \cite{wiechers2011}). However, the attack model in Sec.~\ref{sec:performance} shows that suppression of more pulses requires a higher multiplication factor $x$ to maintain the count rate at Bob, causing the attack to become more difficult for Eve.

\subsection{Exploiting edge-triggered alert monitoring}
\label{sec:trigger_hacking}

\begin{figure}
\begin{subfigure}[t]{\columnwidth}
\includegraphics[width=\columnwidth]{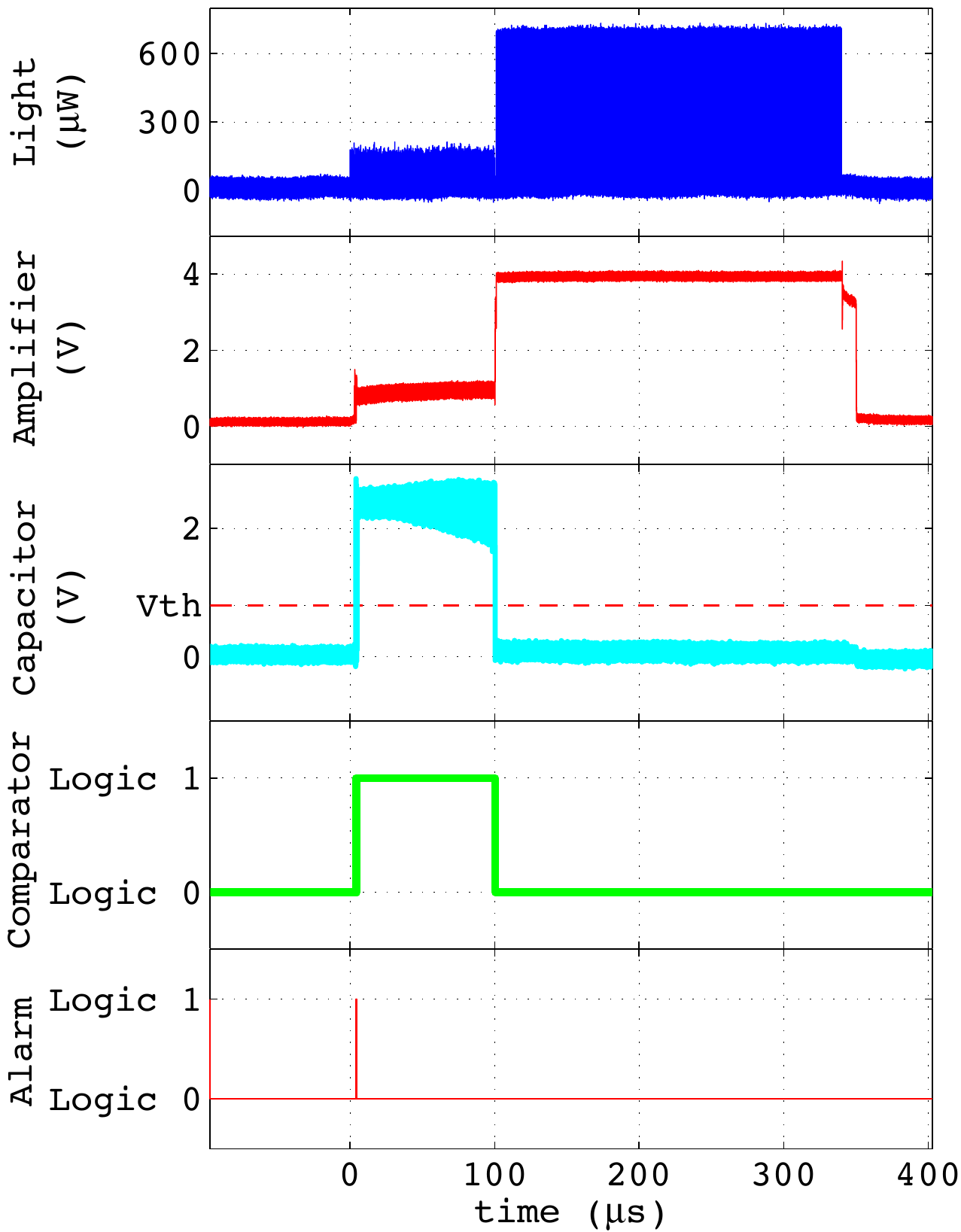}
\caption{Injection of very bright pulses and their effect on the circuit. Note that at the end of the frame when the amplifier output became zero, the capacitor voltage was still low. The reason is because after the end of the frame, the FPGA no longer generated the reset signal and hence the integrator did not reset. It reset at the beginning of the next frame after the reset signal was produced.}
\label{fig:et_whole}
\end{subfigure}
\begin{subfigure}[t]{\columnwidth}
\includegraphics[width=0.72\columnwidth]{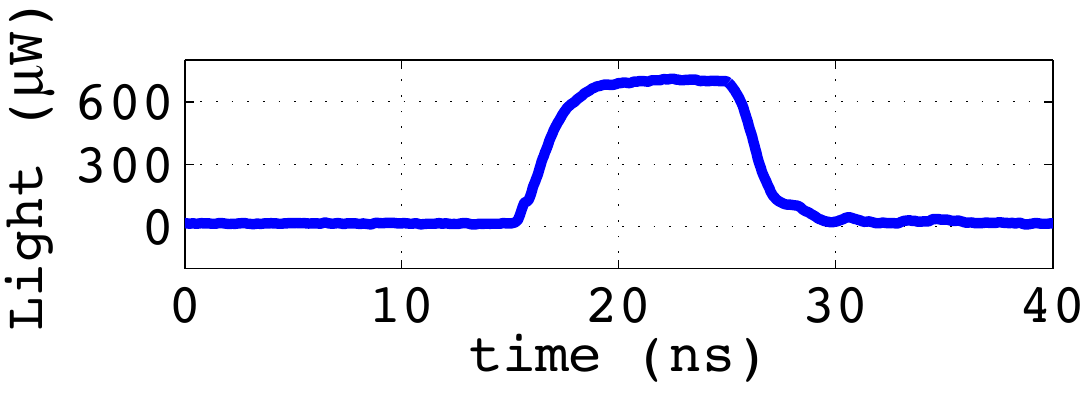}
\caption{Pulse shape carrying the maximum injected energy using this method ($7150\,\femto\joule$ or $97$ times more than the calibrated signal pulse).}
\label{fig:energy_et}
\end{subfigure}
\caption{(Color online) Exploiting edge-triggered alarm monitoring.}
\label{fig:brute}
\end{figure}

As mentioned in Sec.~\ref{sec:pulse_energy_monitoring}, the output from the comparator is applied to a pulse generator that produces a fixed-width alarm pulse on the low-to-high transition of its input. In addition, the integrator is unable to reset the capacitor voltage if the amplifier output is high. These particular design choices pose the biggest loophole in the system, which we have confirmed experimentally. Before the start of the monitoring period, at around $100\,\micro\second$, we started injecting bright probe pulses at each slot in order to push the capacitor voltage completely below the threshold (Fig.~\ref{fig:brute}). As long as the bright pulses were sent (in our case until the end of the frame), the comparator output remained low and there was no low-to-high transition for the pulse generator to produce the alarm. After the end of the frame, when we stopped sending the bright pulses, the amplifier output went low as seen from Fig.~\ref{fig:et_whole} but the capacitor voltage was still below the threshold as there was no reset signal to reset the integrator at the end of the frame. Using this method, we were able to inject probe pulses with a maximum energy of $7150\,\femto\joule$ (limited by our available source power) corresponding to a multiplication factor $x = 97$ (Fig.~\ref{fig:energy_et}). Note that the attack takes place in every bit slot, and no pulses needed to be blocked. Intuitively, at such a high $\mu$ this attack shifts Alice's operation close to a classical regime, and no security can be maintained.

\section{Theory of attacks}
\label{sec:theory}

Clavis2 implements two QKD protocols: non-decoy BB84 and SARG04. In this section we consider BB84, while SARG04 is introduced in Sec. \ref{sec:attack-on-BB84}. The amount of privacy amplification used by Clavis2 to ensure security is based on a strong attack that combines two attacks: photon-number-splitting (PNS) and cloning \cite{curty2004.PhysRevA-69-042321}. With these attacks, the mutual information between Alice and Eve becomes \cite{niederberger2005}
\begin{equation}
I_{A:E} = \frac{1}{2}\mu \eta (t t_b - \frac{\mu}{2})I_1(D_1)+\frac{1}{2}\mu \eta\frac{\mu}{2},
\label{I_AE}
\end{equation}
where $\mu$ is the average photon number per pulse set by Alice, $\eta$ is Bob's average detector efficiency, $t$ is the measured channel transmission efficiency, $t_b$ is the transmission in Bob's interferometer, and $I_1(D_1)$ is the information gathered by Eve when she performed cloning attack  that introduces a disturbance $D_1$ on the state. The first term in the equation comes from the cloning attack, where Eve obtains partial information, and the second term comes from the PNS attack which gives Eve full information. To maintain security, the information gathered by Eve must be removed from the final key:
\begin{equation}
S=I_{A:B}-I_{A:E},
\label{S}
\end{equation}
with $I_{A:B}$ (the mutual information between Alice and Bob) defined as \cite{niederberger2005}: 
\begin{equation}
I_{A:B} = \frac{1}{2}[\mu t t_b\eta +2p_d][1-f_{ec}H(Q)].
\label{I_AB}
\end{equation}
Here, $p_d$ is Bob's detector dark count probability, $f_{ec}$ is the error correction efficiency, $H$ is the binary entropy function, and $Q$ is the measured QBER. The term $f_{ec}H(Q)$ accounts for the information revealed during error correction, which must be discarded. 

The above security analysis makes three basic assumptions. The first is that Eve has no control over Bob's detectors ($\eta$ and $p_d$ cannot be changed). The second is that Bob expects a certain count rate and Eve should not change it. The last assumption is that Eve performs individual attacks. In addition, the analysis ignores multi-photon events above two photons by assuming they occur too infrequently to contribute significantly. Our attack, considered below, follows these three assumptions but includes multi-photon events which become significant as $\mu$ is increased. We also assume that Bob does not monitor double clicks, and instead implements the squashing model \cite{gittsovich2014.PhysRevA-89-012325} (implemented by ID~Quantique in a recent software update to Clavis2), where double clicks are assigned a random bit value, therefore contributing to an average $50\%$ QBER. 

We consider two attacks. The first is a strong attack which is limited by the laws of quantum mechanics only. The second attack is a realistic attack that is limited by the present-day technology.

\subsection{Strong attack}

We model our strong attack as the same combined PNS and cloning attacks assumed by Clavis2 \cite{niederberger2005}, but with $\mu$ being manipulated and increased by a factor $x$ so that the multi-photon components can no longer be ignored. The mutual information between Alice and Eve then becomes:
\begin{equation}
I_{A:E}' = R_1I_1(D_1)+R_{multi}.
\label{I_AE_prime}
\end{equation}
Here $R_1$ ($R_{multi}$) is the contribution to Bob's detection rate from the single-photon (multi-photon) pulses where Eve implements the cloning (PNS) attack.
\begin{equation}
R_1 = \frac{1}{2}p_{attack}^1\eta x \mu e^{-x\mu}
\label{R_1}
\end{equation}
\begin{equation}
R_{multi} = \frac{1}{2}\sum\limits_{n=2}^\infty p_{attack}^n[1-(1-\eta)^{n-1}]\frac{(x\mu)^n}{n!}e^{-x\mu},
\label{R_multi}
\end{equation}
with $p_{attack}^n$ the probability of performing the attack on the n-photon pulse. In cases where Eve doesn't attack on a pulse, this pulse is blocked by her and does not contribute to Bob's detection rates. To ensure the expected count rate at Bob remains unchanged, the rates must follow
\begin{equation}
R_1+R_{multi} = \frac{1}{2}[1-\sum\limits_{n=1}^{\infty}((1-t t_b\eta)^n)\frac{\mu^n}{n!}e^{-\mu}].
\label{R_condition}
\end{equation}

When $x$ is small, $p_{attack}^n$ is always $1$ for $n\ge2$. As $x$ increases, the probability of cloning attacks ($p_{attack}^1$) decreases. If $x$ is large enough for Eq.~(\ref{R_condition}) to be satisfied with $p_{attack}^1=0$, Eve stops performing cloning attacks and begins blocking the pulses with lower photon number to satisfy Eq.~(\ref{R_condition}), i.e.,\ first $p_{attack}^2$ is reduced, then $p_{attack}^3$ and so on until the equation is satisfied.

\subsection{Realistic attack}

Eve's realistic attack is limited by current technologies. In a realistic attack, Eve cannot alter the transmission of the channel, the alignment of the system or characteristics of Bob's detectors. In addition, she must use realistic beamsplitters and optical switches that have non-zero insertion loss.

Eve's realistic attack strategy is to implement an unambiguous state discrimination (USD) attack \cite{dusek2000} with a certain probability $p_{attack}^{USD}$ while doing nothing with a probability $(1-p_{attack}^{USD})$. We also analysed the beam-splitting attack strategy \cite{bennett1992b, Felix2001}, but it performed significantly worse than the USD attack. Hence we only present here the results from the USD attack. In addition, the USD attack has the advantage of producing no extra errors (which could be monitored and used to detect Eve).

\begin{figure}
\includegraphics[width=1.00\linewidth]{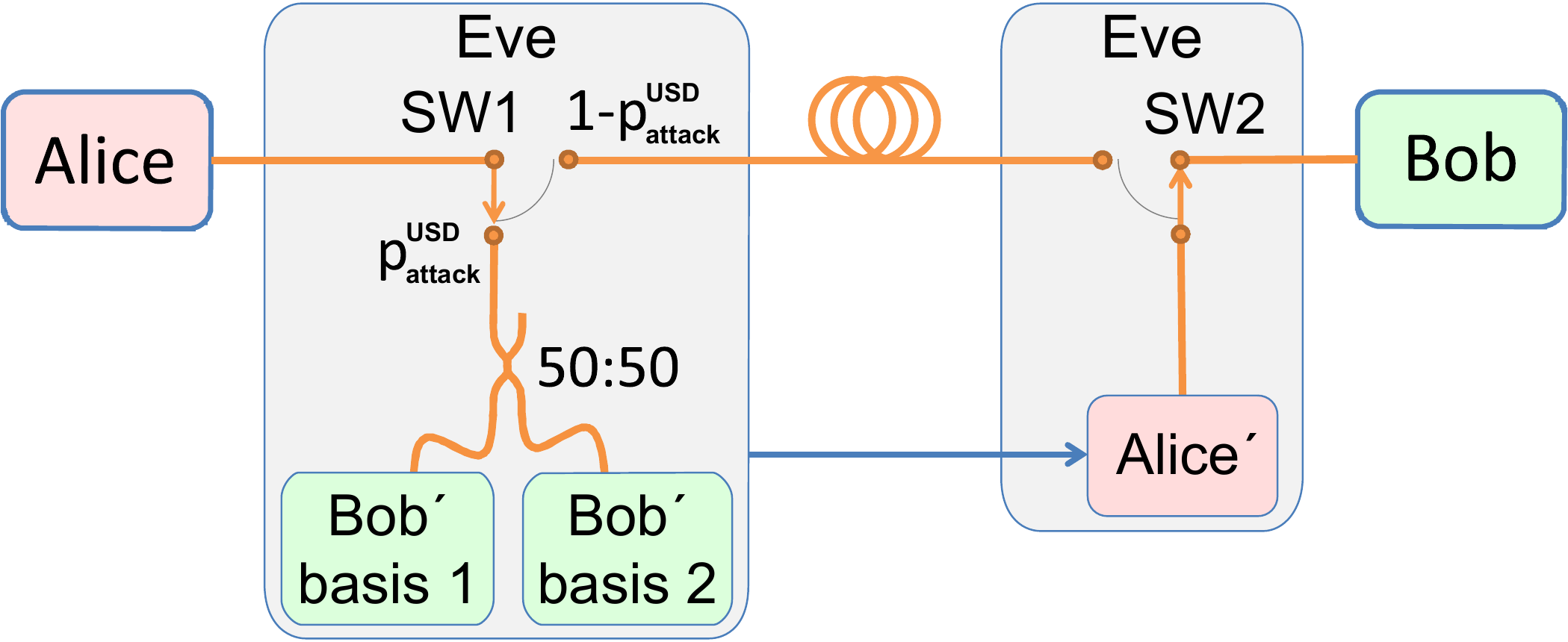}
\caption{(Color online) Realistic attack scheme. With a probability $p_{attack}^{USD}$, pulses from Alice are measured by Eve using a 50:50 beamsplitter followed by two copies of Bob's setup Bob$'$ that use different measurement bases. When the USD measurement is successful, Eve sends a pulse in the measured state using a source Alice$'$ placed next to Bob. SW1 and SW2 are optical switches. SW2 can in practice be replaced by an asymmetric beamsplitter.}
\label{fig:realistic_attack}
\end{figure}

Eve's measurement apparatus, shown in Fig.~\ref{fig:realistic_attack}, consists of a 50:50 beamsplitter followed by two receiver units Bob$'$ (one for each measurement basis) with two detectors each. We assume Eve is placed immediately outside Alice's system (before any transmission losses in the fiber) as this gives Eve the maximum detection probabilities. Eve also has a source Alice$'$, placed just before Bob. This source emits attenuated-laser quantum states with an average photon number $\mu_e$. Using this source, Eve sends a pulse whenever her detections allow her to unambiguously discriminate the state (i.e.,\ when she measures photons in three different detectors, ensuring the correct state is the one measured in the basis with only one detector click). When the state discrimination is ambiguous (measurement in only one or two detectors), she sends nothing. We assume that Eve's alignment is as good as Alice's and Bob's (same fringe visibility $V$), giving Eve's QBER \cite{niederberger2005}
\begin{equation}
Q_e = \frac{1}{2}\left(1-\frac{V}{1+4p_e/(\mu t_{s} t_{BS}\eta_e)}\right),
\label{Q_e}
\end{equation}
where $t_{BS}$ ($t_s$) is the insertion loss of Eve's imperfect beamsplitter (optical switch), $\eta_e$ is the total detection efficiency of Bob$'$ (including its internal losses), and $p_e$ is the detector dark count probability in Bob$'$. The mutual information between Alice and Eve is then:
\begin{equation}
I_{A:E}'' = R_{USD}(1-H(Q_e)),
\label{I_AE_prime2}
\end{equation}
where $R_{USD}$ is the contribution to Bob's detection rate when Eve successfully performs the USD attack. The rate is given by the probability that Eve's measurement is unambiguous multiplied by the probability that Bob registers a measurement in the right basis:
\begin{align}
\begin{split}
R_{USD}=p_{USD}\frac{1}{2}(2p_d+1-e^{-\mu_e t_s t_b\eta}),
\label{R_USD}
\end{split}
\end{align}
where $p_{USD}$ is the probability of an unambiguous measurement by Eve given by the probability of three-detector click:
\begin{equation}
p_{USD}=(1-e^{-x\mu t_{BS} t_s\eta_e/2})(1-e^{-x\mu t_{BS} t_s\eta_e/4})^2
\label{p_USD}
\end{equation}
(see also endnote \footnote{A more general form of $p_{USD}$ can be derived by representing the input state as a sum of Fock states, giving $P_{USD}=\sum\limits_{n=1}^{\infty}p_D^{(n)}p_e^{(n)}$, where $p_D^{(n)}$ is the probability of unambiguous state discrimination when detecting $n$ photons and $p_e^{(n)}$ is the probability of $n$ photons being absorbed by Eve's detectors ($p_e^{(n)}=e^{-x\mu t_{BS} t_s\eta_e}(x\mu t_{BS} t_s\eta_e)^n/n!$ for a weak coherent state). Note that Eve does not require photon number resolving detectors. For our case, where Eve uses a 50:50 beamsplitter to measures in the same two bases as Bob, $p_D^{(n)}=(4^n-2(3^n)+2)/4^n$. This equation is obtained by taking all  $4^n$ possible measurement combinations, subtracting $2(3^n-1)$ combinations that do not yield unambiguous discrimination, then normalizing by dividing by the total $4^n$ combinations. Although the resulting equation for $P_{USD}$ looks different from Eq.~(\ref{p_USD}), we have verified that both forms lead to the same simulated attack performance.}).

In order for Eve to not be detected, she must maintain the expected rate at Bob:
\begin{align}
\begin{split}
p_{attack}^{USD}R_{USD}+(1-p_{attack}^{USD})\frac{1}{2}(2p_d+1-e^{-x\mu t_s^2 t t_b\eta})\\ = \frac{1}{2}(2p_d+1-e^{-\mu t t_b\eta}).
\label{R_condition2}
\end{split}
\end{align}
As $x$ increases, $p_{attack}^{USD}$ will increase, allowing Eve to perform her attack more often. If $x$ is large enough, Eve can perform the attack on every pulse ($p_{attack}^{USD}=1$) without reducing the rate, giving her maximum information.

\section{Performance of attacks}
\label{sec:performance}

\subsection{Assumptions}
\label{sec:Assumptions}

We modeled our attacks using parameters extracted from experimental runs of the Clavis2 system. For several values of channel transmission $t$ we extracted QBER $Q$, fringe visibility $V$, average photon number at Alice's output $\mu$, Bob's detector efficiency $\eta$ and dark count rate $p_d$ (averaged between Bob's two detectors). We used the factory-calibrated value for Bob's interferometer short-arm transmission $t_b$. The number of data pulses sent by Alice was extrapolated based on the number of detections at Bob, $t$, $\mu$, $t_b$, $\eta$ and $p_d$, allowing us to ignore detector deadtime by giving us a number of pulses for which Bob's detectors were sensitive.

Both of our attacks follow the three basic assumptions described in Section \ref{sec:theory}. In the strong attack, Eve uses lossless lines, perfect efficiency detector with no dark counts and perfect alignment, and has access to perfect-efficiency quantum memory and the quantum non-demolition photon-number measurement. For our modeling of the realistic attack, we assume commercially available fiber beamsplitters that can achieve insertion loss as low as $0.3\,\deci\bel$ \cite{Agiltron-coupler} (in addition to splitting loss), and optical switches which can achieve insertion loss of $<1\,\deci\bel$ \cite{BATiSwitch,2Agiltron-switch}. The best detectors that would currently be available for Eve are superconducting nanowire detectors, which are commercially available and have shown both high efficiency ($>90\%$) and very low dark count rate ($<100\,\text{s}^{-1}$) \cite{marsili2013.NatPhotonics-7-210,PhotonSpot}. We assume the total detection efficiency of Bob$'$ $\eta_e = 80\%$, to further account for minor losses in his optical scheme.  We measured the QBER of our Clavis2 system without Eve (for example, in BB84 at $3.4\,\deci\bel$ line loss, it was 1.34\%). In both of our attacks, this measured QBER is used as the minimum QBER for Bob. We allow Eve to increase the QBER in the strong attack to a maximum of 8\%, which is near the limit where Clavis2 can (sometimes) extract secure key \cite{jain2011}. The realistic attack does not cause any increase in QBER because Eve will block all pulses where she does not unambiguously determine the state.

Of the three attacks presented in this paper, the first two require Eve to suppress a certain number of pulses. This limits the information that Eve can gather because she has to maintain the rate at Bob by decreasing the probability of her attack ($p_{attack}^{USD}$). In the third attack, Eve can increase the energy of all pulses, which allows her to get the most information. We used numerical simulation to compute the performance of the attacks.

\subsection{Attack on BB84}
\label{sec:attack-on-BB84}

\begin{figure}
\includegraphics[width=\linewidth]{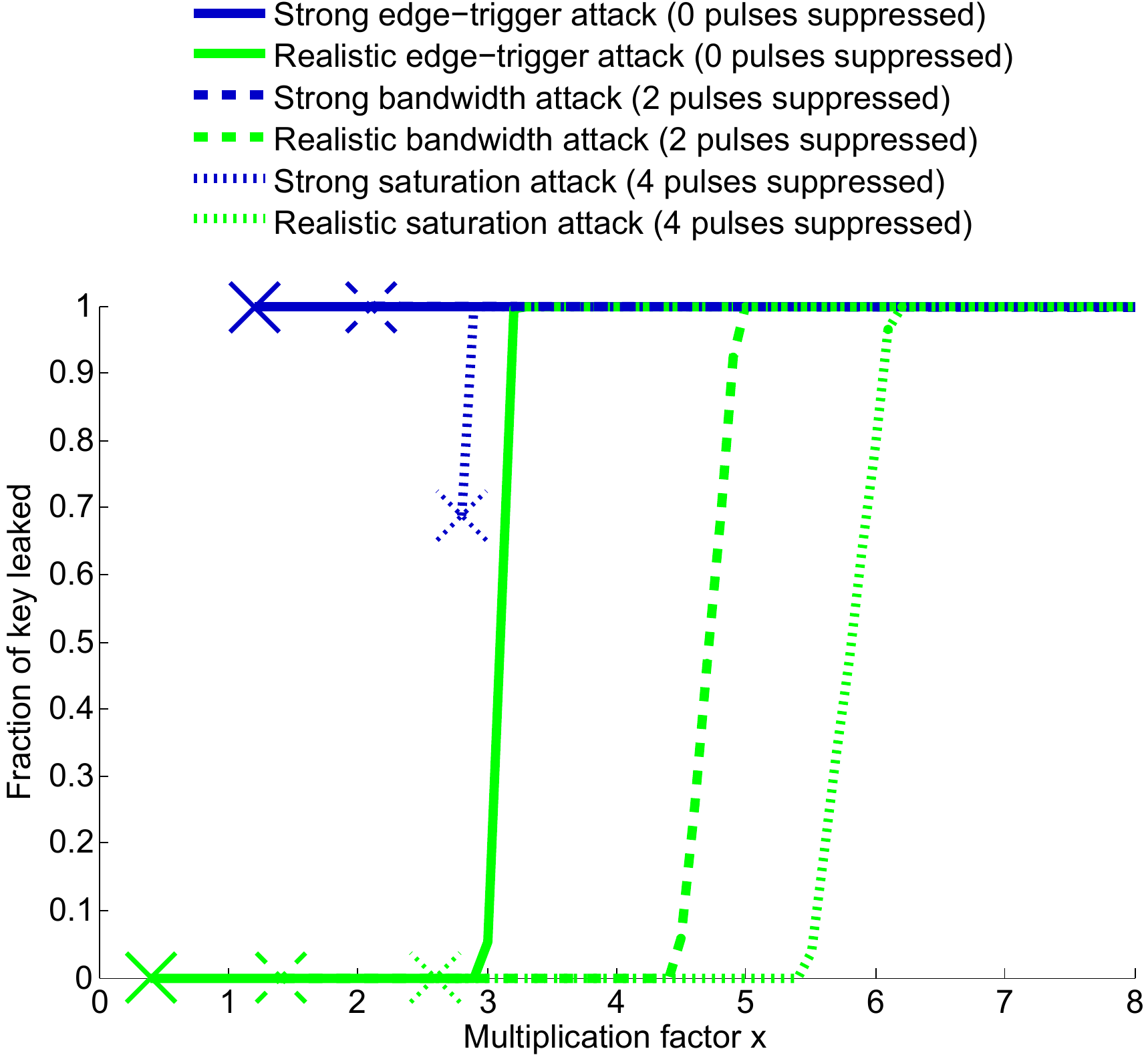}
\caption{(Color online) Fraction of secret key leaked to Eve in the BB84 protocol. The edge-trigger attack, which can increase $\mu$ in all pulses, allows Eve to gain full information with lower multiplication factor $x$ than the attacks that require suppression of pulses. At low $x$ (where the curve stops, marked by the crosses), Eve is unable to maintain the expected count rate at Bob (in the realistic attack), or induces too high QBER (in the strong attack), resulting in her presence being noticed and the key aborted. When the ratio is 0 (realistic attack), Eve is able to maintain the rate but cannot extract sufficient information to overcome privacy amplification. Channel loss is $3.4\,\deci\bel$ and, in the strong attack model, Eve is restricted to a maximum QBER of $8\%$ to avoid suspicion. This maximum QBER value was chosen because it is near the limit where Clavis2 can (sometimes) extract secure key \cite{jain2011}.}
\label{fig:attack_BB84}
\end{figure}

The fraction of secret key that can be known to Eve with the attacks is shown in Fig.~\ref{fig:attack_BB84}. In the bandwidth and saturation attacks, Eve must increase $\mu$ sufficiently to compensate for the suppressed pulses before the attack can be performed without Eve being notice. The bandwidth attack on Clavis2 can increase $\mu$ by up to a factor $x=7.3$ while suppressing two pulses, more than the required $x=5$ to extract full information in the realistic attack model. The performance of the saturation attack is hindered by the large number of pulses suppressed. Nevertheless, the required $x=6.2$ to extract full information in the realistic attack model can be achieved since suppressing four pulses allows $x=7.87$. Both attacks are able to extract full information using the strong attack model, with the bandwidth attack requiring $x=2.7$, while the saturation attacks requires $x=3.8$. The edge-trigger attack, where no pulses are suppressed, allows Eve to extract information with a lower $\mu$ (starting at $x=3$ in the realistic attack model), and can extract full information at $x\ge3.2$. The strong attack is able to extract full information at $x=1.5$.

\begin{figure}
\includegraphics[width=\linewidth]{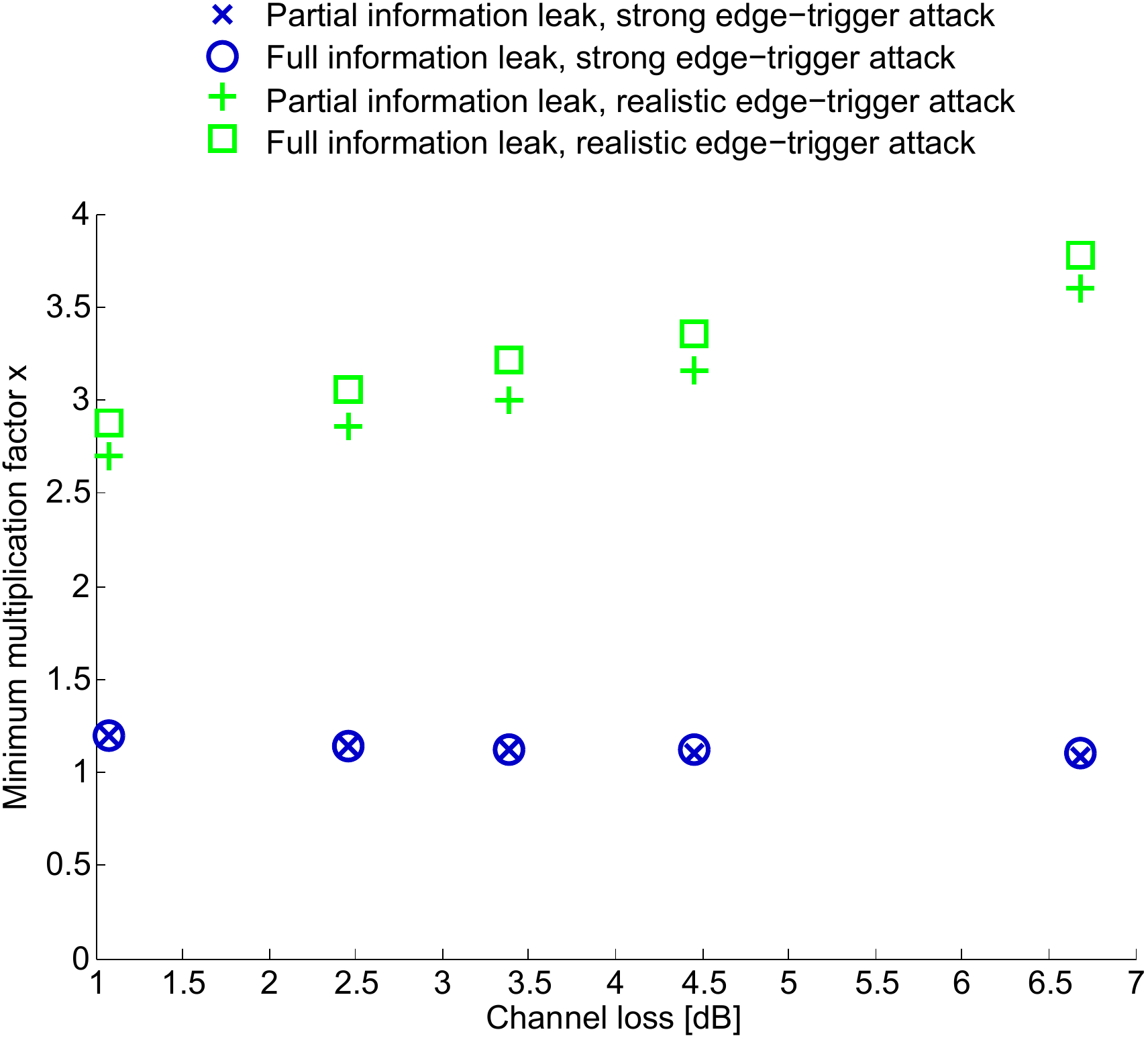}
\caption{(Color online) Minimum $x$ to obtain partial and full information on the secret key in the edge-trigger attack (i.e.,\ with no pulses suppressed) on the BB84 protocol. For the strong (realistic) attack model, Eve is able to extract partial information when between the saltire (cross) and circle (square), and full information above the circle (square). Again, Eve is restricted to a maximum QBER of 8\% to avoid suspicion.}
\label{fig:attack_BB84_vs_loss_no_suppressed_pulses}
\end{figure}

Figure~\ref{fig:attack_BB84_vs_loss_no_suppressed_pulses} shows the dependence of $x$ on channel loss for both partial and full information leak in the edge-trigger attack. The value of $\mu$ depends on the channel loss ($\mu\approx t$ \cite{niederberger2005}), resulting in attack thresholds that only weakly depend on the channel loss, as seen in Fig.~\ref{fig:attack_BB84_vs_loss_no_suppressed_pulses}. Note that commercial Clavis2 systems are only able to extract secure keys up to a certain line loss, limited by detector dark counts. BB84 protocol is more sensitive to loss than SARG04. Our system sample was able to produce secure key with BB84 at up to $6.7\,\deci\bel$ line loss. Beyond this loss, BB84 was never able to extract secure key and thus there was no key information for Eve to gain.

While we have analysed the basic BB84 protocol as implemented in Clavis2, analysis of its detector-decoy \cite{moroder2009} and source-decoy \cite{hwang2003,ma2005} variants can be a future study. In the latter case it is intuitively clear that with a sufficiently high $x$, Eve can distinguish between different decoy states.

\subsection{Attack on SARG04}
\label{sec:attack-on-SARG04}

\begin{figure}
\includegraphics[width=\linewidth]{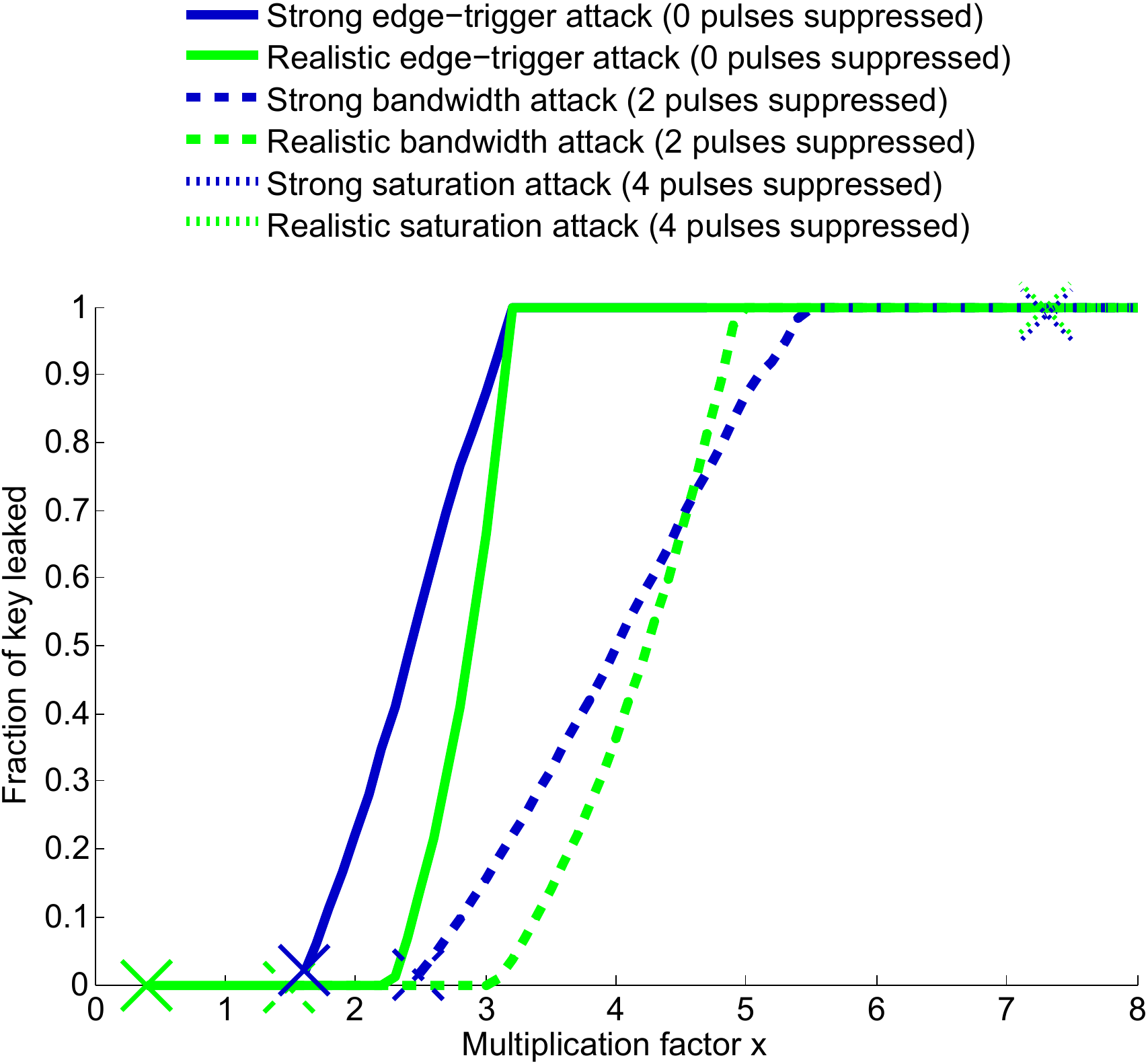}
\caption{(Color online) Fraction of secret key leaked to Eve in the SARG04 protocol. As with BB84, the attack that does not require Eve to suppress pulses performs better than the attacks that require pulse suppression.
% redundant with main text: For strong attacks, the slope of the curve changes around leaked key fraction of 0.35 because Eve begins to block two-photon pulses. A second change occurs above 0.8 where Eve blocks all one-photon and two-photon pulses, relying on only PNS attacks on pulses with three or more photons. 
Once again, the missing points in the curve at low $x$ (marked by the crosses) occur when Eve is unable to maintain the expected count rate at Bob (in the realistic attack), or induces too high QBER (in the strong attack), resulting in her presence being noticed and the key aborted. When the ratio is 0 (realistic attack), Eve is able to maintain the rate but cannot extract sufficient information to overcome privacy amplification. Channel loss is $3.4\,\deci\bel$ and, in the strong attack model, Eve is restricted to a maximum QBER of 8\% to avoid suspicion.}
\label{fig:attack_SARG04}
\end{figure}

In the SARG04 protocol \cite{scarani2004}, keys are encoded in the basis instead of in the state. This lowers the sifting factor to $1/4$ (from BB84's $1/2$) but makes the protocol more robust to PNS attacks. To properly identify the encoded bit, Eve's measurement must return the same outcome as Bob's measurement. Each photon measured by Eve thus has a probability $1/4$ of giving the desired outcome. The probability that Eve fails to gain the right information when measuring $n$ photons is then 
\begin{equation}
E_n = \left(\frac{3}{4}\right)^n.
\label{E_n_SARG04}
\end{equation}
In addition, because the basis is never revealed in the analysis, Eve gains no advantage in waiting until sifting to perform her measurement. We extended both the strong and the realistic attack models to this protocol using Eve's modified probability of failure. The results are shown in Fig.~\ref{fig:attack_SARG04}.

While SARG04 is more resistant to the PNS attack than BB84, it's also less resistant to the USD attack. This is because the SARG04 protocol performs privacy amplification based on the photon-number-splitting attack in which, for one measured photon, Eve extracts only $1/4$ of the information. In comparison, Eve could extract full information in BB84 for one photon measurement using photon-number-splitting attack. However, the information extracted by the USD attack is the same for both SARG04 and BB84, allowing partial key extraction at lower $x$ owing to the reduced privacy amplification performed by the SARG04 protocol. As with BB84, the attacks requiring fewer blocked pulses perform better.

\section{Attack on quantum coin-tossing}
\label{sec:coin-tossing}

Quantum coin tossing (QCT) allows two distrustful parties (Alice and Bob) that are separated by distance to agree on a bit value, while providing security guarantees that are stronger than classically possible. Loss-tolerant strong QCT protocol was first proposed in \cite{berlin2009} and implemented in \cite{berlin2011} with the use of an entangled source. The protocol was slightly modified in \cite{pappa2011} to account for noise in the system, and enabled the implementation of QCT using a plug-and-play system \cite{pappa2014.NatCommun-5-3717}. The two implementations \cite{berlin2011,pappa2014.NatCommun-5-3717} expanded the applicability of quantum information processing beyond QKD. Their results confirmed that using today's technology, QCT can provide a lower cheating probability than its classical counterpart. In this section we demonstrate how a deviation of $\mu$ from the ideal value can affect the performance of the QCT protocol presented in \cite{pappa2011}. In order to take into account all standard experimental imperfections, including channel noise, multiphoton pulses, losses and dark counts, Pappa {\it et al.}\ introduced an honest abort probability $H$, which is the probability that the protocol is unsuccessful when both parties are honest. For a desirable value of $H$, the two players can agree on the value of the protocol parameters, namely the number of protocol rounds $K$, the mean photon number $\mu$ and the state coefficient $y$ of the (rotated) Bell states used by the protocol \cite{pappa2014.NatCommun-5-3717}.

Alice's cheating probability only depends on the coefficient $y$ of the quantum states, therefore a deviation of the mean photon number will not improve her strategy. However, Bob's cheating probability is a function of $\mu$ and can be upper-bounded \cite{pappa2011,pappa2014.NatCommun-5-3717}
\begin{equation}
p^{B}_{cheat} \leq \sum_{i = 1}^{4} P(A_i) P(\text{cheat}|A_i) + [1- \sum_{i =1}^4 P(A_i)].
\label{bob_cheat}
\end{equation}
Here, $P(A_i)$ (for $i=1,\dots,4$) is the probability of the four possible events where Bob receives at most one two-photon pulse in the $K$ protocol rounds, and $P(\text{cheat}|A_i)$ is the maximum cheating probability given that event $A_i$ occurred. For the remaining events, we consider that the cheating probability is $1$ (see the supplementary material of \cite{pappa2014.NatCommun-5-3717} for a more detailed explanation).

\begin{figure}
\includegraphics[width=0.99\linewidth]{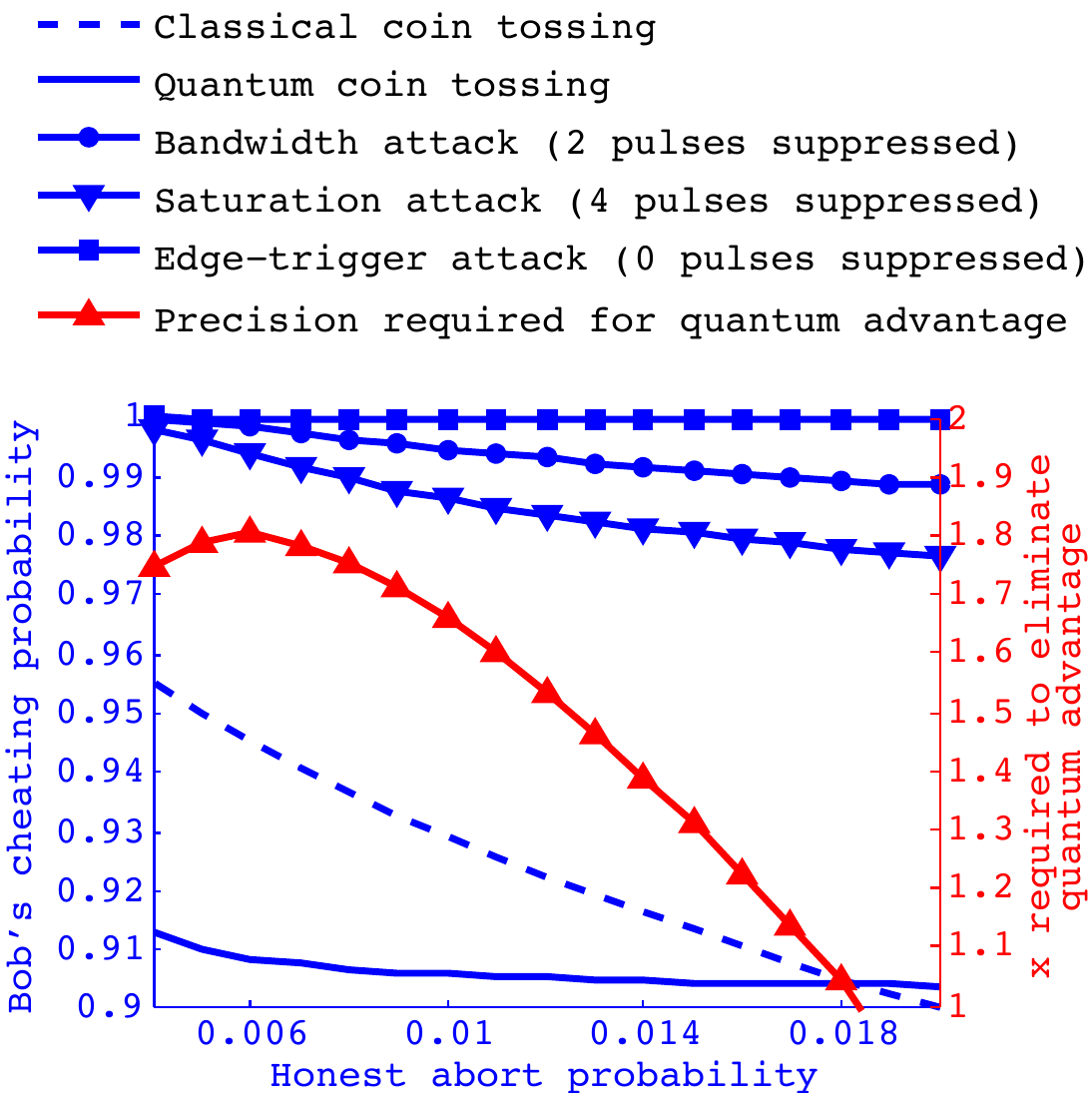}
\caption{(Color online) Bob's cheating probability versus honest abort probability in the coin-tossing protocol. The plot shows limits for the classical coin-tossing and QCT (for $15\,\kilo\meter$ and $\mu=0.0019$ \cite{pappa2014.NatCommun-5-3717}), as well as limits for the three attacks on QCT. We observe that all three Bob's attacks beat the classical limit, and QCT can therefore provide no provable advantage. Also plotted is factor $x$ required to reduce security of QCT to that of its classical counterpart.}
\label{fig:coin_tossing}
\end{figure}

We use the data obtained from the plug-and-play implementation of QCT over $15\,\kilo\meter$ of optical fiber using Clavis2 \cite{pappa2014.NatCommun-5-3717}, to demonstrate how a malicious Bob, having the ability to increase $\mu$ by a factor $x$ without being detected, can increase his cheating probability.  In Fig.~\ref{fig:coin_tossing}, we show the effect of the three attacks presented in this paper, on Bob's cheating probability in comparison with the ideal case where $\mu$ does not deviate from its ideal value (in this case $\mu = 0.0019$) \footnote{Note that in \cite{pappa2014.NatCommun-5-3717} the authors also considered errors during the state preparation (Alice), the choice of measurement basis and bit value, as well as differences in detector efficiencies (Bob). For simplicity, here we assume uniform distribution for Alice's state preparation and Bob's basis, bit choice, as well as equal detector efficiencies}. Using the bandwidth attack for the two-pulse blocking case, the mean photon number increases to $7.3 \, \mu$ while the protocol rounds decrease to $K/3$. For the saturation attack with four-pulse blocking, we have mean photon number $7.87 \, \mu$ and  rounds $K/5$. Finally, for the edge-triggered attack we have used $x = 10$ while keeping the number of protocol rounds the same (i.e.,\ no pulses suppressed), resulting in unity Bob's cheating probability. We note that our modeling here upper-bounds Bob's cheating probability, considering that he has perfect equipment, controls the losses of the channel, and also has the ability to perform quantum non-demolition measurements.

We observe that, if Bob uses any of the three attacks to increase the mean photon number, and is not detected by Alice's pulse-energy-monitoring system, then there is no provable quantum advantage for coin tossing. This means that, similar to QKD, QCT is also vulnerable against the inability to maintain a constant mean photon number. In Fig.~\ref{fig:coin_tossing}, we also show how much manipulation of $\mu$  is required from a quantum Bob in order to increase his cheating probability to the classical limit. For example, for honest abort probability $0.014$, if Bob is able to increase $\mu$ by $x = 1.389$ from the ideal value of $0.0019$, then his cheating probability becomes the same as the classical cheating probability. Equivalently, this means that for this specific honest abort probability, Alice needs to have a measurement precision of $38.9\%$ on the value of $\mu$, if she wants to make her protocol at least as secure as its classical counterpart. So, even if measures are taken to prevent an adversary from manipulating $\mu$, limited experimental precision for setting the exact security parameters inherently affects the protocol performance, and can even make it insecure.

\section{Countermeasures}
\label{sec:countermeasures}

Although the implemented strategy of the pulse-energy-monitoring module is generally correct, the technical realization should be revised dramatically in order to be efficient against arbitrary Trojan-horse attacks. It requires changes in many parts of the circuit: the front-end amplifier, the integrator and the alarm detector.

The negative saturation of the transimpedance amplifier OPA380 can be prevented by pulling down its output by a $2\,\kilo\ohm$ resistor to the $-5\,\volt$ power supply, as advised in the datasheet of the opamp \cite{OPA380}. Nevertheless, the amplifier bandwidth choice, which has been made on a specification considering limited classes of attacks, is not sufficient for the accurate metering of the calibrated signal (second) pulse when Eve can transfer optical energy from the first pulse to the second one. To obtain precision of, say, 10\%, the amplifier output after the first pulse needs to decay to 5\% of its maximum value, since the first pulse is about twice as large as the second pulse. It limits the time constant of an amplifier by the value of $50\,\nano\second / (-\!\ln(0.05)) = 16.7\,\nano\second$, which corresponds to bandwidth of at least $9.5\,\mega\hertz$ (assuming amplifier's frequency response equivalent to an RC-filter). Hence, the front-end transimpedance amplifier should be remodeled to enhance the bandwidth.

At the moment, the integrator circuit functions more like a peak detector than an ideal integrator. Square-law dependence of the FET1 source current on the gate voltage results in non-linearity. It appears that the circuit output is more sensitive to a higher level of the input signal, which is typical for peak-detecting. This way, the circuit actually measures the pulse peak intensity rather than the pulse energy. For proper implementation, the integrator should be built in such a way that the capacitor is charged by current linearly depending on the input voltage.

The edge-triggered alarm generation by means of a monostable is not needed in this circuit at all. Instead, a simple level triggering can be used. Actually, there is no risk of the FPGA missing a too-short electrical pulse at the output of the comparator, because the voltage at C cannot rise until it is reset by the FPGA through FET2. Hence, the monostable can be simply excluded, with possibly slightly delaying resetting the integrator capacitor C to ensure a minimum time to keep the comparator output in a low logic-level state.

Implementing a precise high-speed analog integrator could be challenging. Alternatively the amplifier signal could be digitized with a fast analog-to-digital converter, and the rest of processing done numerically in the FPGA.

The continuous detector is not needed for security if the pulse-energy-monitoring detector is properly implemented. ID~Quantique has been informed about our results prior to this publication, and is developing countermeasures for their affected QKD system.

\section{Conclusion}
\label{sec:conclusion}

In this work, we point out the risk to security that exists when the communicating parties do not have an exact estimate of the system's security parameters ($\mu$ in this case). We also discuss technical measures that should allow to calibrate $\mu$ with an acceptable precision and restore security. Let us remark however that ensuring an accurate knowledge of the security parameters at the time of system installation may not be sufficient. The calibration of equipment may be lost later in the system lifetime either because of a laser-damage attack by Eve \cite{bugge2014}, or because the equipment parameters drift beyond their initial specifications as electronic components age and begin to fail. Tackling this problem is an open question. One possible way to achieve a better understanding is to examine implementation details, no matter how little, more closely than has been done before.

This work also highlights the limitations of closed security standards developed inside a manufacturing company. Although the company in this case went above and beyond everyone else's prior research in this field in order to secure their commercial system (as mentioned in Sec.~\ref{sec:countermeasure-design}), this was not sufficient. In this case, as well as in numerous other instances \cite{lutkenhaus1999,makarov2006,lo2007,zhao2008,lydersen2010a,wiechers2011,sun2011,jain2011,ferenczi2012}, an independent research team uncovered security problems that the original developers of the systems missed. To address this situation, we suggest a two-fold solution. First, open standards on secure implementation and testing of QKD should be developed in a collaboration between the research community and industry. This process is already taking place \cite{langer2009}, but can be intensified in the security specifications aspect. Second, practice shows that independent researchers are usually better at finding security problems than the developers. We therefore think that testing for both unexpected security problems, and for standards compliance, should be led by independent security certification labs.

\begin{acknowledgments}
We thank N.~L{\" u}tkenhaus and N.~Jain for discussions. This work was supported by Industry Canada, Canada Foundation for Innovation, Natural Sciences and Engineering Research Council of Canada, U.S.\ Office of Naval Research, and ID~Quantique. S.S.,\ S.K.\ and J.-P.B\ acknowledge support from CryptoWorks21. S.K.\ acknowledges support from Mike \& Ophelia Lazaridis Fellowship. J.-P.B.\ acknowledges support from FedDev Ontario and Ontario Research Fund. ID~Quantique acknowledges support from European Commission FET QICT SIQS project.
\end{acknowledgments}

\end{document}